\documentclass[%
 aip,
 amsmath,amssymb,
 reprint,%
]{revtex4-1}

\usepackage{graphicx}
\usepackage{dcolumn}
\usepackage{bm}

\usepackage[utf8]{inputenc}
\usepackage[T1]{fontenc}
\usepackage{mathptmx}
\usepackage{etoolbox}

\usepackage{amssymb}
\usepackage{amsmath}
\usepackage{amsthm}
\usepackage{color}
\usepackage{comment}
\usepackage{enumitem}
\usepackage{graphicx}
\usepackage{listings}
\usepackage{xcolor}
\usepackage{caption}

\theoremstyle{thmstyleone}%
\theoremstyle{thmstyletwo}%

\theoremstyle{thmstylethree}%
\newtheorem{definition}{Definition}%

\makeatletter
\def\@email#1#2{%
 \endgroup
 \patchcmd{\titleblock@produce}
  {\frontmatter@RRAPformat}
  {\frontmatter@RRAPformat{\produce@RRAP{*#1\href{mailto:#2}{#2}}}\frontmatter@RRAPformat}
  {}{}
}%
\makeatother
\begin{document}

\preprint{AIP/123-QED}

\title[On the higher-order smallest ring-star network of Chialvo neurons under diffusive couplings]{}
\author{Anjana S.~Nair}
\affiliation{School of Digital Sciences, 
Digital University Kerala, 
Technopark Phase-IV campus, Mangalapuram - 695317,
Kerala, India}

\author{Indranil Ghosh}%
\altaffiliation[\textbf{Author to whom correspondence should be addressed:}]{ i.ghosh@massey.ac.nz}
 \email{i.ghosh@massey.ac.nz}
\affiliation{ 
School of Mathematical and Computational Sciences,
Massey University,
Colombo Road, Palmerston North-4410,
New Zealand
}%

\author{Hammed O.~Fatoyinbo}
\affiliation{%
Department of Mathematical Sciences, School of Engineering, Computer and Mathematical Sciences, Auckland University of Technology, Auckland-1142, New Zealand
}%

\author{Sishu S.~Muni}
\affiliation{School of Digital Sciences, 
Digital University Kerala, 
Technopark Phase-IV campus, Mangalapuram - 695317,
Kerala, India}

\date{\today}

\begin{abstract}
We put forward the dynamical study of a novel higher-order small network of Chialvo neurons arranged in a ring-star topology, with the neurons interacting via linear diffusive couplings. This model is perceived to imitate the nonlinear dynamical properties exhibited by a realistic nervous system where the neurons transfer information through higher-order multi-body interactions. We first analyze our model using the tools from nonlinear dynamics literature: fixed point analysis, Jacobian matrix, and bifurcation patterns. We observe the coexistence of chaotic attractors, and also an intriguing route to chaos starting from a fixed point, to period-doubling, to cyclic quasiperiodic closed invariant curves, to ultimately chaos. We numerically observe the existence of codimension-1 bifurcation patterns: saddle-node, period-doubling, and Neimark Sacker. We also qualitatively study the typical phase portraits of the system and numerically quantify chaos and complexity using the 0-1 test and sample entropy measure respectively. Finally, we study the collective behavior of the neurons in terms of two synchronization measures: the cross-correlation coefficient, and the Kuramoto order parameter. 
\end{abstract}

\maketitle

\begin{quotation}
Recently, the application of higher-order interactions in network neurodynamics has been a topic of conversation. This study answers whether higher-order interaction among neurons in a small network promotes chaos and synchronization. The small network under consideration is a ring-star network (made to be a topologically complete graph) of Chialvo neurons, with one node at the center and three at the periphery, linked via diffusive couplings (both pairwise and higher-order). We delve into the dynamical properties of this small network via fixed point analysis, bifurcation patterns, and time-series analysis, before concluding with a report on the synchronization behavior of the nodes acting in a collective manner in the network. 
\end{quotation}

\section{Introduction}
Dynamical properties of complex systems is a prominent branch of computational science, that has garnered the attention of physicists, mathematical modelers, quantitative biologists, computer scientists, neuroscientists, etc. {\em Neurodynamics} is a subfield that deals with the dynamical properties of neurons~\cite{Iz07, GeKi14}, usually studied via the mathematical tools like ordinary differential equations, maps, partial differential equations, and delay differential equations. Thus, nonlinear equations are commonly used to illustrate the evolution of the dynamic firing and bursting behaviors exhibited by the neurons~\cite{HeYa21}. Some popular neurodynamical models studied quite intensively are the Fitzhugh-Nagumo model~\cite{Fi61}, the Hindmarsh-Rose model~\cite{HiRo84}, and the Hodgkin-Huxley model~\cite{HoHu52}. Usually, the dynamics are studied on a neuron, or a collection of multiple neurons forming a complex topological structure. To model these complex structures, mathematical modelers utilize the concept of {\em networks}, where the nodes are perceived as neurons and the edges are as electrical or chemical synaptic connections among the neurons. Then the behavior is studied over time, which not only depends on the entity concerned but also on the other entities it is connected to~\cite{GaDi21}. Usually, in network dynamics, most studies are based on the assumption that all interactions are dyadic or pairwise (only two nodes are involved in the interaction)~\cite{GaGh23, GhVe23}. However, in many real-life scenarios, interactions are not pairwise but higher-order, involving more than two nodes. Some application areas of higher order networks are in the spread of rumors, peer pressure in social systems, disease spread, and the competency for food by multiple species in a complex ecosystem comprising non-pairwise interactions (see the review report by Battiston {\em et al.}~\cite{BaCe20}). Another application area is the molecular network modeling protein chemical structures and pathways~\cite{GaMa18}. There is no doubt that higher-order interaction models in networks have found a well-deserved place in the neurodynamics literature~\cite{YuYa11, InMo09, PaMe22, MiMe22, NoGa20, GaCo21}. Higher-order networks in neurodynamics have found applications in neurodegenration~\cite{HeRo22,PrVi03}, and many psychiatric disorders~\cite{LeLe23, SiAn24} Interested readers with a physics background can have a look at the work by Battiston {\em et al.}~\cite{BaAm21}, where the physics of higher-order interactions in complex systems are discussed.

Our work focuses on a ``small'' network. A small network is a minimal model made up of three to ten coupled oscillators working collectively as a unit. Our work is motivated by the broad need to study complex collective groups in terms of reduced mathematical models. The nervous system serves as an ideal candidate to be studied on the basis of minimal models, for example, the smallest network possible for a particular topology. This is because the small network acts as a unit of a larger ensemble replicating the behavior of a real-world nervous system. Another advantage of studying small networks is they fill the gap between studies concerning a single oscillator model and multi-oscillator models consisting of at least a hundred oscillators. Some of the simple topologies that have proliferated the network neurodynamics literature are the ring network~\cite{OmLa22, KhPa19}, star network~\cite{YaMa22, RaPy21}, ring-star network~\cite{MuPr20, MuFa22, GhMu23}, lattice network~\cite{VaWa05, PaMo14}, and multiplex network~\cite{LeYa21, WeWu15}. Motived, we introduce the smallest network of neuron oscillators (consisting of four oscillators), which are arranged in a ring-star topology: one in the center and the rest in the periphery.

Most studies in neurodynamics are based on continuous-time ordinary differential equations~\cite{Da12} with discrete-time systems comparatively less studied. However, discrete-time models help in the faster, simpler, more flexible, and computationally more optimal understanding of neural dynamics\cite{IbCa11}. They also provide insights into how the neurons depend on parameters such as refractory period\cite{BeMe98}, firing threshold, and network architecture. Some of the discrete neuron models studied are Chialvo\cite{Ch95}, Rulkov\cite{LiBa22}, Izhikevich\cite{Va10}, and discrete Hindmarsh Rose model\cite{HiRo84}. The focus of our work is the Chialvo neuron, where each oscillator in our novel network follows the dynamics set by the Chialvo map, first put forward by Dante Chialvo in 1995. In the Chialvo neuron model, the slowness of the recovery variable is not constrained, so this model exhibits a wide range of dynamical behaviors, such as subthreshold oscillations, bistability, and chaotic orbits.\cite{IbCa11}. Studies on the ring-star network of Chialvo neurons have been reported recently by Muni {\em et al.}~\cite{MuFa22}, and Ghosh {\em et al.}~\cite{GhMu23}, but both these studies were considered in the thermodynamic limit, i.e, a large number of oscillators. Also, an additional topic in these two studies was the application of electromagnetic flux. Furthermore, no higher-order interactions were considered between the oscillators. In this manuscript, however, the difference from those articles is three-fold:
\begin{enumerate}
    \item We consider the smallest network of ring-star topology with just four nodes,
    \item no electromagnetic flux is applied, and
    \item the interaction among the nodes is beyond pairwise, i.e., higher-order.
\end{enumerate}

For the parameter values that have been investigated, our model exhibits various dynamical patterns like fixed points, period doubling, cyclic quasiperiodic closed invariant curves, and chaos. We also report the phenomenon of coexistence, where for a particular set of parameter values, we see the existence of attractors which can be both periodic or chaotic. This is known as {\em multistability}~\cite{Fe08} and has been achieved by utilizing the tool of bifurcation analysis. Many real-world systems exhibit the multistability phenomenon, which appears unsuitable in engineering systems because multiple states cause errors in the system, but it is a useful phenomenon in neural systems that process information\cite{MuFa22}. Moreover, the appearance of regular and chaotic dynamics makes it obligatory to distinguish them mathematically to obtain a more accurate understanding of the behavior of the system~\cite{ChSa90}. For this, a variety of methods such as the $0-1$ test~\cite{GoMe09, GoMe16} and the standard Lyapunov exponent are commonly used in numerics. Nevertheless, it might not be possible to use the conventional Lyapunov exponent method for higher-dimensional systems~\cite{Po13}. Thus, we implement the $0-1$ test to the time series data generated from our system for differentiating between the chaotic and regular behavior portrayed. Additionally, the abundant complexity of the network calls for a measure to quantify it and we do that via an {\em entropy} measure called the {\em sample entropy} first introduced by Richman {\em et al.}~\cite{RiMo00} where they measured the complexity in physiological time series data. We have also applied sample entropy to the time series data of our system. The concept of entropy in connection with the information theory was first introduced by Shannon \cite{Sh01}, and this concept has subsequently been employed in various studies related to neuroscience~\cite{TiLa18, FaDe23, ViWi11}. Richman and his coauthors' method is by far one of the most popular entropy measures for nonlinear dynamical analysis other than the {\em approximate entropy} devised by Pincus~\cite{Pi91}. In terms of applications, measuring the complexity of empirical data from EEG recordings can point towards a possible arrival of Alzheimer's disease~\cite{WaZh19}. Performing time series analysis by implementing the $0-1$ test and the sample entropy measure on our time series provides a plausible numerical bridge between chaoticity and complexity in our model.

Furthermore, as our model is a collection of oscillators connected to one another forming a group, it becomes necessary to study what kind of synchronization patterns this network generates. Synchronization is an important feature of neuron ensembles and has been extensively studied in the field of dynamical systems. This phenomenon is visible at a wide array of spatiotemporal scales~\cite{BlHu99, Co18}, and has been explored in various fields of study. A list of important reads are Sumpter~\cite{Su06}, Strogatz~\cite{St03}, and Shahal {\em et al.}~\cite{ShWu20}. Synchronization can give rise to both normal and abnormal patterns. Kuramoto~\cite{KuBa02} was the one who first identified a {\em chimera} state and it refers to the coexistence of both coherent and incoherent nodes within a certain ensemble of coupled phase oscillators. A handful of nodes will oscillate away from the main synchronized ensemble in this case, giving rise to a chimera. Also, sometimes the nodes oscillate in two or more subgroups, giving rise to cluster states. These states have been observed in various natural phenomena, such as the sleep patterns of animals~\cite{MaLe05}, the flashing patterns of fireflies~\cite{SaPe22}, and many other systems~\cite{HaSc15, MaTh13, WiKi13}. To quantify synchronization, we have utilized two metrics, the {\em cross-correlation coefficient}~\cite{VaSt16, SeVa18} and the {\em Kuramoto order parameter}~\cite{Ku84, St00, BiGo20}. Both of these have found a wide range of applications in the study of network dynamical systems~\cite{ShMu21, RySc23, Pr23, AnPr22}.

We organize this article as follows: In section~\ref{sec:hon} we discuss what a higher-order network is, and put forward some basic mathematical notations related to hypergraphs and simplicial complexes, that make up higher-order networks. In section~\ref{sec:Chialvo}, we review the two-dimensional Chialvo map, that acts as the building block of our network model. Then in section~\ref{sec:network}, we introduce our novel higher-order neuron network which is built on the smallest ring-star topology consisting of four Chialvo nodes, with linear diffusive couplings. We also comment on the topology of this network in terms of graph theory terms like the adjacency matrices. We further delve into analyzing the fixed point of this system, write down the Jacobian of the system, and explore its eigenvalues at the fixed point in section~\ref{sec:fp}. This sets up an overview of the basic dynamical properties our system exhibits. In section~\ref{sec:bif} we show the bifurcation diagram of the action potentials of the neurons with respect to the higher-order coupling strength (primary bifurcation parameter). We observe an interesting route to chaos: the appearance of fixed points, period doubling, cyclic quasiperiodic closed invariant curves, and finally chaos. Additionally, we implement \textsc{MatContM} in section~\ref{sec:matcontm} to report more codimension-$1$ bifurcation patterns (saddle-node, and Neimark-Sacker) numerically, utilizing \textsc{MatContM}.  These bifurcation patterns appear in our model on the variation of the higher-order coupling strength in the negative domain, which we otherwise do not observe if not for \textsc{MatContM}. Next, in section~\ref{sec:tsa}, we first look into the typical phase portraits that appear at certain interest values of the higher order coupling strength, before implementing the $0-1$ test and sample entropy measure on our model and numerically test for chaos and complexity. Finally, in section~\ref{sec:sync}, we employ the synchronization measures, the cross-correlation coefficient, and the Kuramoto order parameter on our model. We provide concluding remarks and future directions in section~\ref{sec:conc}. We have performed all the numerical simulations with \texttt{Python 3.9.7} extensively using \texttt{numpy}, \texttt{pandas}, and \texttt{matplotlib}, except for \textsc{MatContM}~\cite{MeGo17, KuMe19} which is originally based on \texttt{MATLAB}.

\section{Higher Order Network}
\label{sec:hon}
In real-world systems, interactions can occur as both pairwise and non-pairwise in a network. A network can be modeled by a mathematical object called a {\em graph} which consists of nodes and edges. They can reveal information about the dynamical and structural characteristics of a system. Most often in graph-based methods, relationships between entities (via the edges) are considered as pairwise or dyadic. In real-life scenarios, however, nondyadic network relations and interactions involving the nonlinear interaction of more than two nodes are more often. To acquire insights from such complex systems, it is important to understand and analyze the polyadic network interactions. Such networks are called higher-order networks consisting of higher-order interactions. These interactions can be mathematically encoded through {\em hypergraphs} and {\em simplicial complexes}. A hypergraph is a generalized version of a graph where edges can connect any number of vertices greater than or equal to two. We first set up some definitions of hypergraphs and simplicial complexes following Bick {\em et al.}~\cite{BiGr23}
\begin{definition}
    For a finite set of vertices $\mathcal{V}$ and its collection of nonempty subsets $\mathcal{L} \subseteq \mathcal{P}(\mathcal{V})$, where $\mathcal{P}(\mathcal{V})$ is the power set of $\mathcal{V}$, the tuple $\mathcal{H} = \left(\mathcal{V}, \mathcal{L} \right)$ is called a hyprgraph.
\end{definition}
Like graphs, a hypergraph can be weighted and serves as the main backbone for the model of this paper. A good resource to study more about hypergraphs is the book by Berge~\cite{Be84}.

\begin{definition}
Let $\mathcal{V} = \left\{1, \ldots, M \right\}$, where $M$ is the number of vertices. Then an $s$-simplex $\tau$ is 
\begin{enumerate}
    \item $\tau = \left\{v_0, \ldots, v_s \right\}$ is a set of $s+1$ elements of $\mathcal{V}$,
    \item $\tau \ne \varnothing$, and
    \item $\tau$ has dimension $s$.
\end{enumerate}
\end{definition}

\begin{definition}
    Given an $s$-simplex $\tau$, a face $\mathcal{F}$ of $\tau$ is a subset $\mathcal{F} \subset \tau$.
\end{definition}

\begin{definition}
    A simplicial complex $\mathcal{C}$ is a collection of a finite number of $s-$simplices having the following properties
    \begin{enumerate}
        \item $\mathcal{F} \subset \tau$ and $\tau \in \mathcal{C}$ imply that $\mathcal{F}$ is an element of $\mathcal{C}$, and
        \item for all vertices $v \in \mathcal{V}$, $\left\{v \right\} \in \mathcal{C}$.
    \end{enumerate}
\end{definition}

For example, let $\mathcal{V} = \left\{1, 2, 3, 4\right\}$. The maximum possible dimension of the simplices for this case is $3$. The largest possible simplex for this case is the $3$-simplex, given by the quadruple $\left\{1,2,3,4 \right\}$ itself. The next possible simplices are the $2$-simplices represented by the four triples (or triangles) $\left\{1, 2, 3 \right\}$, $\left\{1, 2, 4 \right\}$, $\left\{1, 3, 4 \right\}$, and $\left\{2, 3, 4 \right\}$. Then we have the $1$-simplices represented by the six doubles (sometimes called links) as $\left\{1, 2\right\}$, $\left\{1, 3\right\}$, $\left\{1, 4\right\}$, $\left\{2, 3\right\}$, $\left\{2, 4\right\}$, and $\left\{3, 4\right\}$. Finally, we have the $0$-simplices represented by the four points (or vertices/nodes) as $\left\{ 1\right\}$, $\left\{ 2\right\}$, $\left\{ 3\right\}$, and $\left\{ 4\right\}$. It should be noted that the curly braces ``$\left\{ \right\}$'' represent the {\em roster} notation of a set. In section~\ref{sec:network} we will introduce a higher-order network model with four nodes which will be built on $s$-simplexes, where $s$ will be maximum $2$, and the dynamics of each of these oscillators will be modeled by the Chialvo neuron.

\section{Chialvo neurons}
\label{sec:Chialvo}
In this section, we review the two-dimensional Chilvo map~\cite{Ch95} which is the building block of our network model. Ibarz {\em et al.}~\cite{IbCa11} provides a detailed review on this topic. The Chialvo map is a two-dimensional iterative map characterizing the general dynamics of excitable neuron systems, given by

\begin{align}
\label{eq:Chialvo}
x(n+1) &= x(n)^2e^{(y(n) - x(n))} +k_0 \\
y(n+1) &= ay(n) - bx(n) +c,
\end{align}
where $x$ is the action potential or the activation potential variable and $y$ is the recovery variable at the time step $n$. The four control parameters are $a$, $b$, $c$, and $k_0$, out of which the first two represent the time constant and the activation dependence of recovery respectively, both set to be less than $1$. The variable $c$ is an offset and $k_0$ is an additive perturbation which is time dependent. The excitability of the model is represented by $y$'s fast recovery dynamics. The Chialvo map is able to analyze the transitions from non-autonomous to autonomous behavior, as well as from regular to chaotic patterns, in a small parameter space. As mentioned in Ibarz {\em et al.}~\cite{IbCa11}, the model also exhibits subthreshold oscillations, bistability, and chaotic orbits, making it an ideal candidate for map-based neuron modeling techniques. This has also received quite an attention from the research community in recent times~\cite{WaCa18, PiGr24, MuFa22, GhMu23, XuHu23, BaRy23}.

\section{Network Model}
\label{sec:network}
\begin{figure}[h]
  \centering
  \includegraphics[width=0.5\linewidth]{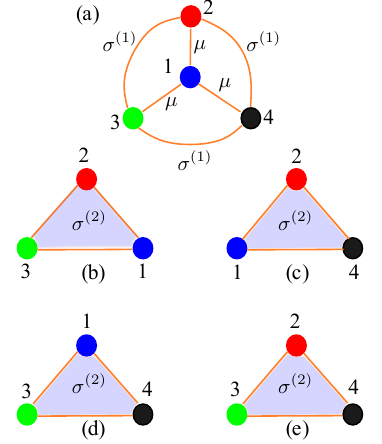}
  \caption{Schematic of the smallest ring-star network of Chialvo neurons with higher-order interactions. The nodes are numbered and color-coded. The coupling strength with the star configuration is denoted by $\mu$, while $\sigma^{(1)}$ represents the coupling strength within the ring. Moreover, $\sigma^{(2)}$ denotes the coupling strength of higher-order interactions for the $2$-simplexes.}
  \label{fig:schematic}
\end{figure}

Our model is the smallest ring-star network made of four oscillators with one at the center and three at the periphery, all linked to each other, forming a complete network (or a complete graph), see Fig.~\ref{fig:schematic}. The dynamics of each of these oscillators is governed by the discrete Chialvo map, discussed in section~\ref{sec:Chialvo}. Note that we incorporate simplicial complexes in our network model to allow for higher-order interactions among the neuron oscillators. This implies a multi-body interlinkage among the oscillators which is beyond dyadic. We allow a maximum of $2$-simplexes within the model to make it have a minimal higher-order relationship, keeping in mind that our model is very minimal. Interested researchers are welcome to try to include a $3$-simplex, which would be possible with our model.

We number the central node as $1$ (colored blue), and the peripheral nodes as $2$ (colored red), $3$ (colored green), and $4$ (colored black). These neuron nodes/oscillators by themselves are the $0$-simplexes. The pairwise couplings for the star configuration, i.e., from node $1$ to $2$, $3$, and $4$ are denoted by $\mu$. The pairwise couplings for the ring configuration, i.e., among the nodes $2$, $3$, and $4$ are denoted by $\sigma^{(1)}$. Thus, in total we have six $1$-simplexes, given by the doubles $\left\{1, 2\right\}$, $\left\{1, 3\right\}$, $\left\{1, 4\right\}$, $\left\{2, 3\right\}$, $\left\{2, 4\right\}$, and $\left\{3, 4\right\}$, see Fig.~\ref{fig:schematic}-(a). Next, we consider the higher-order interactions realized by triadic couplings, i.e., between three oscillators. Our ring-star topology is an ideal candidate to churn out four (which is a very good number for a small network model) such $2$-simplexes given by the triples $\left\{1, 2, 3 \right\}$, $\left\{1, 2, 4 \right\}$, $\left\{1, 3, 4 \right\}$, and $\left\{2, 3, 4 \right\}$. These coupling strengths are denoted by $\sigma^{(2)}$, see Fig.~\ref{fig:schematic}-(b)$\to$(e).

This system of the smallest ring-star higher-order network of Chialvo neurons gives rise to a nonlinear system of eight coupled equations, which in compact form is written as
\begin{align}
    \label{eq:model_p}
        x_p(n+1) &= x_p(n)^2e^{(y_p(n)-x_p(n))} + k_0 + \mu(x_1(n)-x_p(n)) \nonumber \\
        &+ \sigma^{(1)}\sum_{i=2}^4(x_i(n) - x_p(n))\nonumber \\&+\sigma^{(2)}\sum_{i=1}^4\sum_{\substack{j=i+1 \\ i\ne p \\ j\ne p}}^4(x_i(n) + x_j(n) - 2x_p(n)),\\
        y_p(n+1) &= ay_p(n) - bx_p(n)+c,
    \end{align}
    and
    \begin{align}
    \label{eq:model_1}
    x_1(n+1) &= x_1(n)^2e^{(y_1(n) - x_1(n))}+k_0 + \mu\sum_{i=2}^4(x_i(n) - x_1(n)) \nonumber \\
    &+\sigma^{(2)}\sum_{i=2}^4\sum_{\substack{j=i+1}}^4(x_i(n) + x_j(n) - 2x_1(n)), \\
    y_1(n+1) &= ay_1(n) - bx_1(n) + c.
    \end{align}
Observe that for the peripheral nodes, the dynamical behavior is governed by~\eqref{eq:model_p} where $2\le p\le 4$ designates a peripheral node, whereas the dynamical behavior of the central node is governed by~\eqref{eq:model_1}. Moreover, note that we allow the coupling strengths to be both positive and negative since, both excitatory and inhibitory neurons take part in the brain activity~\cite{Li23}. We have followed the simplest diffusive linear couplings within the $s$-simplexes. These kinds biologically interpret electrical synapses between neurons. Similar coupling has been implemented in higher-order Hindmarsh-Rose neurons by Parastesh {\em et al.}~\cite{PaMe22}, along with more complicated nonlinear couplings arising from chemical synapses. Another well-studied neuron map is the Rulkov map, and the higher-order interactions in a network of Rulkov neurons have been reported by Mirzaei {\em et al.}~\cite{MiMe22}. For a detailed review, the readers can refer to Majhi {\em et al.}~\cite{MaPe22}. 

Let the set of dynamical variables at step $n$ be given by
\begin{align}
\label{eq:dynamic_variables}
X = \left\{x_1(n), y_1(n), x_2(n), y_2(n), x_3(n), y_3(n), x_4(n), y_4(n) \right\}.
\end{align}
This network is asymmetric, meaning it changes its form under the transformation $X \to -X$. It is a quenched system, i.e., the dynamics of the model over time are realized on the oscillators, however, the coupling links are static. The topological property of the couplings can be understood mathematically through the adjacency matrices of the network. The adjacency matrix of the pairwise interactions from Fig.~\ref{fig:schematic}-(a) capturing the $1$-simplexes is given by
$$
    \begin{bmatrix}
      0 & \mu & \mu & \mu \\
      \mu &0 & \sigma^{(1)}& \sigma^{(1)} \\
      \mu &\sigma^{(1)} & 0 & \sigma^{(1)} \\
      \mu & \sigma^{(1)} & \sigma^{(1)} & 0
    \end{bmatrix},
$$
whereas the four adjacency matrices for three-body interactions capturing the $2$-simplexes are
$$
    \sigma^{(2)}\begin{bmatrix}
      0 & 1 & 1 & 0 \\
      1 &0 & 1& 0 \\
      1 &1 & 0 & 0 \\
      0 & 0 & 0 & 0
    \end{bmatrix},
    $$
    $$
    \sigma^{(2)}\begin{bmatrix}
      0 & 1 & 0 & 1 \\
      1 &0 & 0& 1 \\
      0 &0 & 0 & 0 \\
      1 & 1 & 0 & 0
    \end{bmatrix},
    $$
    $$
    \sigma^{(2)}\begin{bmatrix}
      0 & 0 & 1 & 1 \\
      0 &0 & 0& 0 \\
      1 &0 & 0 & 1 \\
      1 & 0 & 1 & 0
    \end{bmatrix},
    $$
and
    $$
    \sigma^{(2)}\begin{bmatrix}
      0 & 0 & 0 & 0 \\
      0 &0 & 1& 1 \\
      0 &1 & 0 & 1 \\
      0 & 1 & 1 & 0
    \end{bmatrix},
    $$
for Fig.~\ref{fig:schematic}-(b), (c), (d), and (e) respectively.

The ring-star topology of the network model plays a significant role in capturing the dynamics of three variants of topologies usually studied in the network neurodynamics literature: the ring network, the star network, and the ring-star network. Our model can be reduced to a ring network by setting $\mu=0$ and $\sigma^{(1)} \ne 0$, whereas to a star network by setting $\mu \ne 0$ and $\sigma^{(1)}=0$. However, how to handle $\sigma^{(2)}$ in both these cases, opens a new research avenue. We can think of this model biologically implying a hub-based unit of neurons replicating how the {\em multipolar} neurons form an ensemble for information processing in the {\em central nervous system}. The central node is a hub or the multipolar neuron that transmits information to and from the peripheral multipolar neurons via the dendrites. The diffusive coupling strengths imitate how the dendrites control the information processing via electrical signals distributed through the {\em axons}. The higher-order interactions bring about how these electrical signals are processed among these neurons via more complicated and realistic ways, in terms of mathematical modeling of biological systems. Also, our network model can be perceived as a unit that repeats itself to generate a more complex ensemble of neurons all connected in a complicated topology via diffusive couplings, in the real-world nervous system. Thus studying the dynamical properties of this unit will give a researcher an idea of the bigger picture of the intricate complexity of the nervous system.

Furthermore, our model is map-based, making it computationally efficient, and can be utilized to analyze empirical data from the likes of MRI and EEG. This can be achieved by explaining the firing patterns of the excitable neurons. The study of bifurcation through these modeling techniques reveals how the neurons transition from being at a stable state of some degree to another stable state, with random periods of chaotic firings as certain activities are performed. Looking into the chaotic behavior of these firing patterns also helps researchers identify neural disorders. Identifying disorders is the first step towards treating them and has become an essential field of research to aid medicine~\cite{WaG24}. These modeling techniques via bifurcation patterns pointing towards how the transition happens from stable state to chaotic state is a good reference for understanding a body's circadian rhythms~\cite{HaHa16}. Furthermore, studying how each of these neurons in the network synchronizes with each other over time, tells us how perfectly the information processing occurs among them. A group of neurons are synchronized with another when they organize themselves to form functional ensembles, oscillating in tandem with each other. The global and local communications in an ensemble drive all the sensory, motor, and cognitive tasks~\cite{ScJo05}. When two neurons are synchronized, this indicates the neural activities between them are accurate~\cite{Gr04}. Often an abnormal synchronization refers to the inception of a probable pathophysiological mechanism underlying motor symptoms like Parkinson's disease~\cite{ScJo05, Br03} and Epilepsy~\cite{Ma10}. Thus, these models become quite essential in analyzing empirical data and aiding biomedical sciences.

\section{Fixed point analysis}
\label{sec:fp}
Analyzing the fixed points of a system (continuous or discrete time) is the first step toward unfolding its complex dynamics. The fixed point of a map-based model $f(\mathbf{x})$ is the point $\mathbf{x}^*\in \mathbb{R}^N$, such that $f(\mathbf{x}^*)=\mathbf{x}^*$. Let the fixed point of the system~\eqref{eq:model_p} and~\eqref{eq:model_1} be given by 
\begin{align}
\label{eq:fp}
X^* = \left(x_1^*, y_1^*, x_2^*, y_2^*, x_3^*, y_3^*, x_4^*, y_4^* \right).
\end{align} 
To compute $X^*$, the following list of equations needs to be solved:
\begin{align}
\label{eq:fp_x1}
 x_1^* &= {x_1^*}^2e^{(y_1^*-x_1^*)}+k_0+(\mu+2\sigma^{(2)})(x_2^*+x_3^*+x_4^*-3x_1^*)\\
 \label{eq:fp_y1}
    y_1^* &= ay_1^*-bx_1^*+c\\
    \label{eq:fp_x2}
    x_2^* &= {x_2^*}^2e^{(y_2^*-x_2^*)}+k_0+\mu(x_1^*-x_2^*)+\sigma^{(1)}(x_3^*+x_4^*-2x_2^*)\nonumber \\
    &+2\sigma^{(2)}(x_1^*+x_3^*+x_4^*-3x_2^*)\\
    \label{eq:fp_y2}
    y_2^* &= ay_2^*-bx_2^*+c\\
    \label{eq:fp_x3}
    x_3^* &= {x_3^*}^2e^{(y_3^*-x_3^*)}+k_0+\mu(x_1^*-x_3^*)+\sigma^{(1)}(x_2^*+x_4^*-2x_3^*)\nonumber \\
    &+2\sigma^{(2)}(x_1^*+x_2^*+x_4^*-3x_3^*)\\
    \label{eq:fp_y3}
    y_3^* &= ay_3^*-bx_3^*+c\\
    \label{eq:fp_x4}
    x_4^* &= {x_4^*}^2e^{(y_4^*-x_4^*)}+k_0+\mu(x_1^*-x_4^*)+\sigma^{(1)}(x_2^*+x_3^*-2x_4^*)\nonumber \\
    &+2\sigma^{(2)}(x_1^*+x_2^*+x_3^*-3x_4^*)\\
    \label{eq:fp_y4}
    y_4^* &= ay_4^*-bx_4^*+c   
\end{align}

A step towards gaining this is to try eliminating the $y_p^*$ terms such that we end up with four transcendental equations involving $x_p^*$. From~\eqref{eq:fp_y1} we have
\begin{align}
\label{eq:y1_star}
    y_1^* = \frac{bx_1^* - c}{a-1},
\end{align}
which we substitute in~\eqref{eq:fp_x1} to get
\begin{align}
\label{eq:x1_star}
x_1^* &= {x_1^*}^2e^{\left(\frac{(b-a+1)x_1^*-c}{a-1}\right)}+k_0\nonumber \\
&+(\mu+2\sigma^{(2)})(x_2^*+x_3^*+x_4^*-3x_1^*).
\end{align}
Similarly, we will have
\begin{align}
\label{eq:yp_star}
y_p^* = \frac{bx_p^* - c}{a-1},\qquad p=2,3, 4,
\end{align}
which we substitute in the corresponding $x_p^*$'s to get
\begin{align}
\label{x2_star}
x_2^* &= {x_2^*}^2e^{\left(\frac{(b-a+1)x_2^*-c}{a-1}\right)}+k_0+\mu(x_1^*-x_2^*)\nonumber \\
&+\sigma^{(1)}(x_3^*+x_4^*-2x_2^*)+2\sigma^{(2)}(x_1^*+x_3^*+x_4^*-3x_2^*),\\
\label{x3_star}
x_3^* &= {x_3^*}^2e^{\left(\frac{(b-a+1)x_3^*-c}{a-1}\right)}+k_0+\mu(x_1^*-x_3^*)\nonumber \\
&+\sigma^{(1)}(x_2^*+x_4^*-2x_3^*)+2\sigma^{(2)}(x_1^*+x_2^*+x_4^*-3x_3^*),\\
\label{x4_star}
x_4^* &= {x_4^*}^2e^{\left(\frac{(b-a+1)x_4^*-c}{a-1}\right)}+k_0+\mu(x_1^*-x_4^*)\nonumber \\
&+\sigma^{(1)}(x_2^*+x_3^*-2x_4^*)+2\sigma^{(2)}(x_1^*+x_2^*+x_3^*-3x_4^*). 
\end{align}
We see that~\eqref{eq:fp_x1}, \eqref{eq:fp_x2}--\eqref{eq:fp_x4} are a set of four highly coupled nonlinear transcendental equations which we can numerically solve by using any user-friendly computational solvers, for example, \texttt{fsolve()} function from \texttt{Python}'s \texttt{scipy.optimize} module. Then we substitute these values to give us the $y^*$'s, ultimately fetching us $x^*$'s, eventually giving us the fixed point $X^*$. Now the dynamics of the perturbation vector $X - X^*$ is given by \begin{align}
    \begin{bmatrix}
        x_1(n+1) \\
        y_1(n+1) \\
        x_2(n+1) \\
        y_2(n+1) \\
        x_3(n+1) \\
        y_3(n+1) \\
        x_4(n+1) \\
        y_4(n+1)
    \end{bmatrix} = 
    J.\begin{bmatrix}
        x_1(n) \\
        y_1(n) \\
        x_2(n) \\
        y_2(n) \\
        x_3(n) \\
        y_3(n) \\
        x_4(n) \\
        y_4(n)
    \end{bmatrix},
\end{align} 
where $J$ is the $8 \times 8$ Jacobian matrix, given by
\begin{align}
    J = \begin{bmatrix}
        J_{x_1x_1} & J_{x_1y_1} & J_{x_1x_2} & 0 & J_{x_1x_3} & 0 & J_{x_1x_4} & 0 \\
        -b & a & 0 & 0 & 0 & 0 & 0 & 0 \\
        J_{x_2x_1} & 0 &J_{x_2x_2} & J_{x_2y_2} & J_{x_2x_3} & 0 & J_{x_2x_4} & 0\\
        0 & 0 & -b & a & 0 & 0 & 0 & 0 \\
        J_{x_3x_1} & 0 & J_{x_3x_2} & 0 & J_{x_3x_3} & J_{x_3y_3} & J_{x_3x_4} & 0\\
        0 & 0 & 0 & 0 & -b & a & 0 & 0\\
        J_{x_4x_1} & 0 & J_{x_4x_2} & 0 & J_{x_4x_3} & 0 & J_{x_4x_4} & J_{x_4y_4} \\
        0 & 0 & 0 & 0 & 0 & 0 & -b & a
    \end{bmatrix}, \nonumber
\end{align}
with 
\begin{align}
    J_{x_1x_1} &= x_1(2-x_1)e^{(y_1-x_1)}-3(\mu + 2\sigma^{(2)}), \\
    J_{x_py_p} &= x_p^2e^{(y_p-x_p)},\qquad p = 1,\ldots, 4 \\
    J_{x_1x_2} &= J_{x_1x_3} = J_{x_1x_4} \nonumber \\
    &= J_{x_2x_1} = J_{x_3x_1} =  J_{x_4x_1} = \mu+2\sigma^{(2)}, \\
    J_{x_qx_q} &= x_q(2-x_q)e^{(y_q-x_q)}-(\mu + 2\sigma^{(1)} + 6\sigma^{(2)}),\nonumber \\
    &\qquad q=2, 3, 4 \\
    J_{x_2x_3} &= J_{x_2x_4} = J_{x_3x_2} = J_{x_3x_4} = J_{x_4x_2} \nonumber \\
    &= J_{x_4x_3}= \sigma^{(1)} + 2\sigma^{(2)}.
\end{align}
The absolute values of the eigenvalues of $J$ determine the linear stability analysis of $X^*$. The eight eigenvalues $\lambda_i, i = 1, \ldots, 8$ can be evaluated from $J$ at $X^*$ by solving an eighth order polynomial equation $P_8(\lambda)=0$ where
\begin{align}
\label{eq:poly}
P_8(\lambda)&=a_0\lambda^8+a_1\lambda^7+a_2\lambda^6+a_3\lambda^5+a_4\lambda^4\nonumber \\
&+a_5\lambda^3+a_6\lambda^2+a_7\lambda+a_8.
\end{align}
As this becomes algebraically intractable, one can approach this problem by using any standard computational solver, for example, we have used the \texttt{eig()} function from \texttt{Python}'s \texttt{numpy.linalg} module. If all the eigenvalues $|\lambda_i| < 1$, then $X^*$ is locally stable, and if all $|\lambda_i|>1$, then $X^*$ is locally unstable. Otherwise, $X^*$ is a $k-$saddle where $k$ is the number of eigenvalues whose absolute value is $>1$. 

\begin{figure}[h]
\centering
\begin{tabular}{cc}
  \includegraphics[scale=0.25]{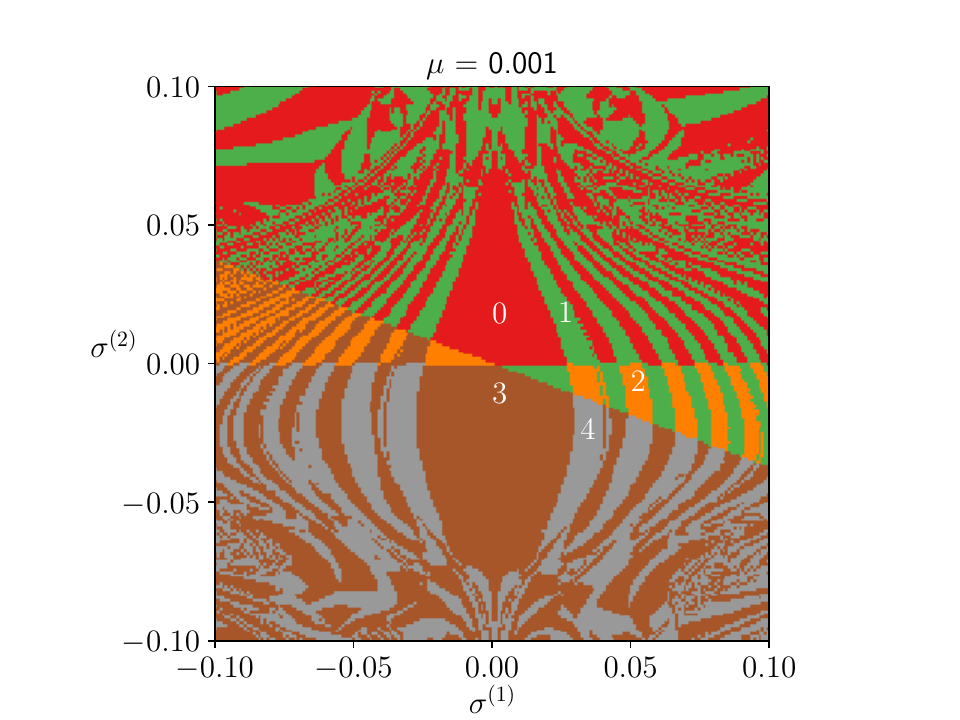} &   \includegraphics[scale=0.25]{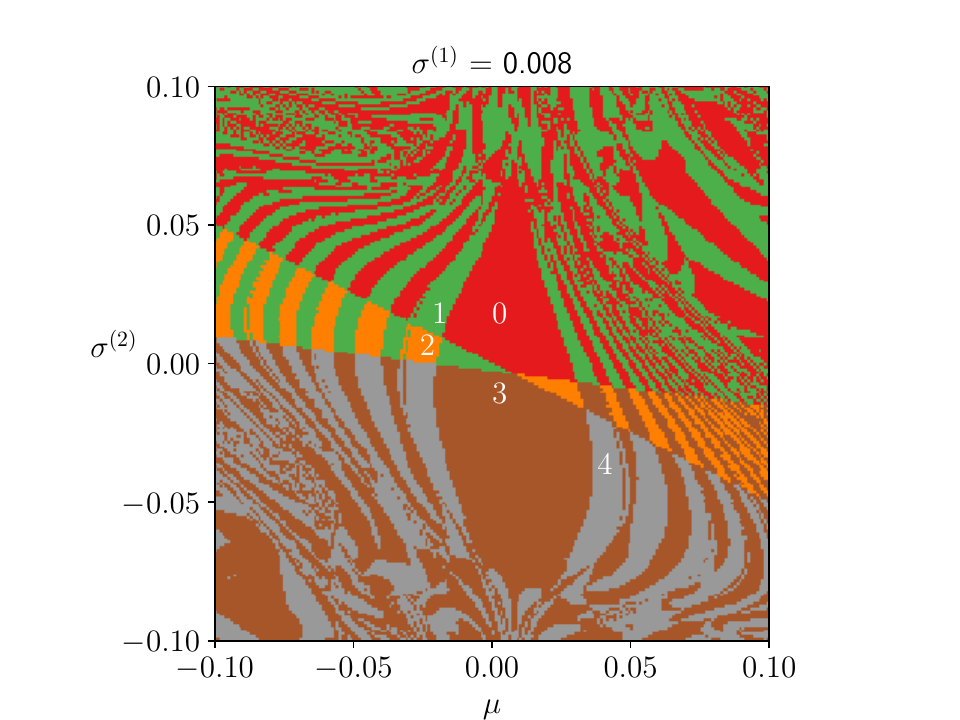} \\
(a) $\sigma^{(1)}$ vs $\sigma^{(2)}$ & (b) $\mu$ vs $\sigma^{(2)}$ \\[3pt]
\end{tabular}
\caption{
Two-dimensional color-coded stability region plots. Both panels show $k-$saddle type and unstable fixed points. Panel (a) shows a $(\sigma^{(1)}, \sigma^{(2)})$ plane with $\mu=0.001$ and panel (b) shows a $(\mu, \sigma^{(2)})$ plane with $\sigma^{(1)}=0.008$. Other parameters are kept as $a=0.759$, $b=0.421$, $c=0.84$, $k_0=0.03$. For the saddle type fixed points, $k$ varies from $1$ to $4$. The stable fixed points are the regions denoted by red pixels, having the number $0$ marked in white. These regions have all eigenvalues $|\lambda_i|<1$ for all $i=1, \ldots, 8$. The other $k$-saddle regions are marked by the number $1\to 4$ in white.
}
\label{fig:stab}
\end{figure}

In Table~\ref{tab:stab}, we list down the set of eight eigenvalues for a parameter set $(\sigma^{(1)}, \sigma^{(2)})$ at $\mu = 0.001$ and the other local parameters set as $a=0.759$, $b=0.421$, $c=0.84$, and $k_0 = 0.03$. We observe a stable fixed point and fixed points of $k$-saddle type with $k=1, \ldots, 4$. Extending on this, it becomes essential to draw a two-dimensional color coded plot, making the higher-order coupling strength $\sigma^{(2)}$ vary in one of the axes. This is like a two-dimensional bifurcation plot, where we focus on the qualitative behavior of the stability of fixed points on the variation of two coupling strengths as bifurcation parameters, see Fig.~\ref{fig:stab}. In panel (a) we vary $\sigma^{(1)}$ against $\sigma^{(2)}$, both in the range $[-0.1, 0.1]$ with $\mu = 0.001$, and in panel (b), we vary $\mu$ against $\sigma^{(2)}$, both in the range $[-0.1, 0.1]$ with $\sigma^{(1)} = 0.008$. We number the regions (in white) according to the stability of the fixed point. Note that the red pixels denote the region containing stable fixed points, where $k=0$. All eigenvalues have absolute values greater than 1. The other pixels are colored according to the $k$-saddle type where $k = 1, \ldots, 4$. These numbers are indicated in white on the plots. We observe distinct bifurcation boundaries where the stability of the fixed point changes from being stable to $k$-saddle types.

\begin{table*}
    \centering
    \begin{tabular}{|c|c|c|c|c|}
    \hline
         $(\sigma^{(1)}, \sigma^{(2)})$& $\lambda_i$ & $\lvert\lambda_i \rvert$ & $\lvert\lambda_i \rvert>1$ & Type \\
         \hline \hline
         $(0.01, 0.02)$ & $0.99987648$ & $0.99987648$ & F & stable \\
         & $0.80477184+0.03787732i$ & $0.80566271$ & F & \\
         & $0.80477184-0.03787732i$ & $0.80566271$ & F&\\
         & $0.77365338$ & $0.77365338$ & F& \\
         & $0.79126146+0.04988789i$ & $0.79283258$ & F& \\
         & $0.79126146-0.04988789i$ & $0.79283258$ & F& \\
         & $0.79126148+0.04988788i$ & $0.7928326$ & F& \\
         & $0.79126148-0.04988788i$ & $0.7928326$ & F& \\
         \hline
         
         $(0.023, 0.02)$ & $1.00027466$ & $1.00027466$ & T & $1$-saddle \\
         & $0.80488425+0.03776246i$ & $0.80576961$ & F & \\
         & $0.80488425-0.03776246i$ & $0.80576961$ & F&\\
         & $0.77364068$ & $0.77364068$ & F& \\
         & $0.77199429+0.05800085i$ & $0.77417006$ & F& \\
         & $0.77199429-0.05800085i$ & $0.77417006$ & F& \\
         & $0.77199427+0.05800085i$ & $0.77417004$ & F& \\
         & $0.77199427-0.05800085i$ & $0.77417004$ & F& \\
         \hline
         
         $(0.05, -0.01)$ & $1.07964252$ & $1.07964252$ & T & $2$-saddle \\
         & $1.00025013$ & $1.00025013$ & T & \\
         & $0.77001084$ & $0.77001084$ & F&\\
         & $0.77364145$ & $0.77364145$ & F& \\
         & $0.9223877$ & $0.9223877$ & F& \\
         & $0.92238763$ & $0.92238763$ & F& \\
         & $0.78062413$ & $0.78062413$ & F& \\
         & $0.78062414$ & $0.78062414$ & F& \\
         \hline

         $(0.01, -0.01)$ & $1.07957293$ & $1.07957293$ & T & $3$-saddle \\
         & $0.99992094$ & $0.99992094$ & F & \\
         & $0.77365195$ & $0.77365195$ & F&\\
         & $0.77001168$ & $0.77001168$ & F& \\
         & $1.05149891$ & $1.05149891$ & T& \\
         & $1.05149886$ & $1.05149886$ & T& \\
         & $0.77106815$ & $0.77106815$ & F& \\
         & $0.77106815$ & $0.77106815$ & F& \\
         \hline

         $(0.033, -0.022)$ & $1.17824731$ & $1.17824731$ & T & $4$-saddle \\
         & $1.00026082$ & $1.00026082$ &T & \\
         & $0.77364111$ & $0.77364111$ & F&\\
         & $0.76742141$ & $0.76742141$ & F& \\
         & $1.08001272$ & $1.08001272$ & T& \\
         & $1.08001265$ & $1.08001265$ & T& \\
         & $0.77000632$ & $0.77000632$ & F& \\
         & $0.77000632$ & $0.77000632$ & F& \\
         \hline
    \end{tabular}
    \caption{Eigenvalues associated with the stability analysis of the fixed points at the parameter point $(\sigma^{(1)}, \sigma^{(2)})$ with $\mu = 0.001$. The local parameters of the system are set as $a=0.759$, $b=0.421$, $c=0.84$, and $k_0 = 0.03$.}
    \label{tab:stab}
\end{table*}

\section{Bifurcation structure}
\label{sec:bif}
Once we have an overall idea about the fixed point of the network system, and its stability regions, the next step is to delve into the bifurcation pattern of its dynamical variables. In this section, we look into how the action potentials of the network change qualitatively over the variation of a bifurcation parameter. We consider the higher-order coupling strength $\sigma^{(2)}$ as the primary bifurcation parameter, and plot $x_1$, $x_2$, $x_3$, and $x_4$ against it. The purpose of these bifurcation patterns is to shed light on the structural stability of the network system and to realize the sensitivity of the action potentials on the primary bifurcation parameter. 

\begin{figure}[h]
  \centering
  \includegraphics[width=0.5\linewidth]{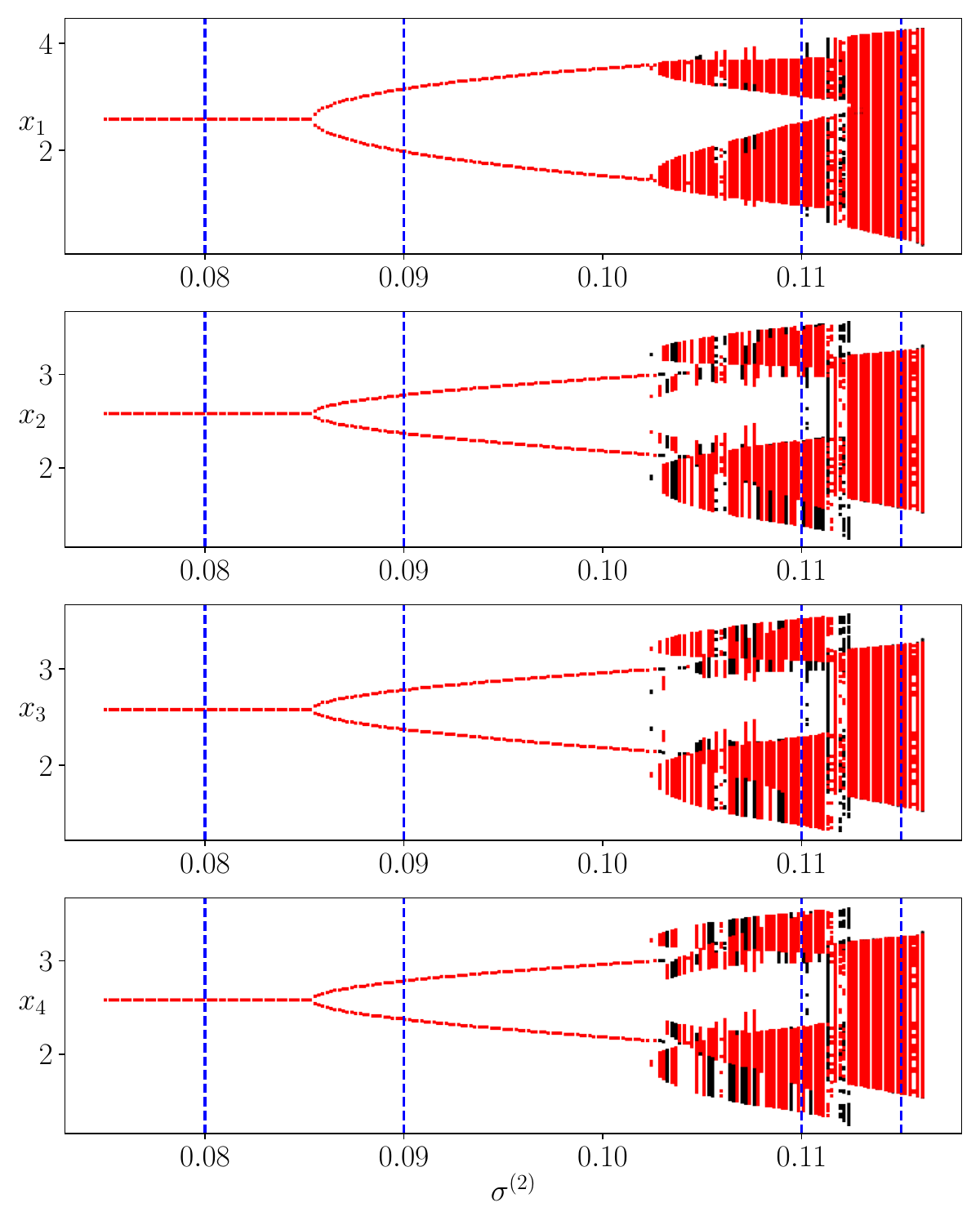}

  \caption{Bifurcation plot of each node against the coupling strength $\sigma^{(2)}$ is plotted, varying $\sigma^{(2)}$ from $0.075$ to $0.116$, once continued forward (points colored black) and once backward (points colored red). The system is iterated for $50000$ steps and only the last $5000$ points are plotted. The initial conditions of the dynamical variables are sampled uniformly from $[0.6, 0.8]$. The parameter values are set as $a=0.89$, $b=0.28$, $c=0.901$, $k_0=0.06$, $\mu=0.03$, and $\sigma^{(1)}=0.001$. It shows the presence of multistability and chaotic attractors. The blue dashed lines indicate the $\sigma^{(2)}$ values for which we explore the dynamics of our model in more detail.}
  \label{fig:bif}
\end{figure}

Figure~\ref{fig:bif} represents the bifurcation plot obtained by plotting each node against the non-pairwise coupling strength $\sigma^{(2)}$. The system is simulated for $5 \times 10^4$ iterations, with the last $5 \times 10^3$ iterations used for plotting to avoid transient effects. We plot $200$ values of $\sigma^{(2)}$ in the range $[0.075, 0.116]$, considering both forward and backward continuation. Forward continuation points are marked in black, whereas the backward continuation points are marked in red. The local parameter values are set as $a=0.89, b=0.28, c=0.901, k_0=0.06, \mu=0.03, \sigma^{(1)}=0.001$. The dynamical variables are initialized by sampling uniformly from the range $[0.6, 0.8]$. As $\sigma^{(2)}$ is varied, we see the appearance of a fixed point (period-$1$, see Fig.~\ref{fig:bif_pp}-(a)) in the range $\sigma^{(2)} \in [0.075, 0.08543]$. At $\sigma^{(2)} \approx 0.08543$, there is a period-doubling (see Fig.~\ref{fig:bif_pp}-(b)), which continues until $\sigma^{(2)} \approx 0.10237$ at which point we observe the onset of quasi-periodicity. On the further increase of $\sigma^{(2)}$, we see the appearance of smooth, disjoint, cyclic, quasiperiodic, closed invariant curves (see Fig.~\ref{fig:bif_pp}-(c)), which merge into single chaotic attractor when $\sigma^{(2)}$ is increased even more via loss of this smoothness through attractor merging crisis. This transition from period-$1$ solution to chaos is discussed in more detail in section~\ref{sec:tsa} with illustration of relevant typical phase portraits and implementing the $0-1$ chaos test on the bifurcation pattern. The reasoning behind continuing the model forward and backward in the same simulation environment is to allow us to report coexistence, indicating {\em multistability}. The attractors are topologically distinct from each other at some $\sigma^{(2)}$ values indicated by nonoverlapping of the black and red marks in Fig.~\ref{fig:bif_pp}. Note that we study the phase portraits at different $\sigma^{(2)}$ values of interest in more detail in section~\ref{sec:tsa}. These are marked as vertical blue lines in Fig.~\ref{fig:bif_pp}, i.e., $\sigma^{(2)}=0.08, 0.09, 0.11, 0.115$.

It is also possible to observe the behavior by varying the pairwise coupling strengths $\mu$ and $\sigma^{(1)}$ and generate two more bifurcation plots for our model. The existence of fixed-point, period doubling, and chaotic behavior can also be observed from these bifurcation plots. Note that we chose not to report these results in this work, focusing more on the higher-order coupling strength as the primary bifurcation parameter. 

\section{Codimension-$1$ bifurcation patterns}
\label{sec:matcontm}
\begin{figure}[!h]
 \centering
 \includegraphics[scale=0.175]{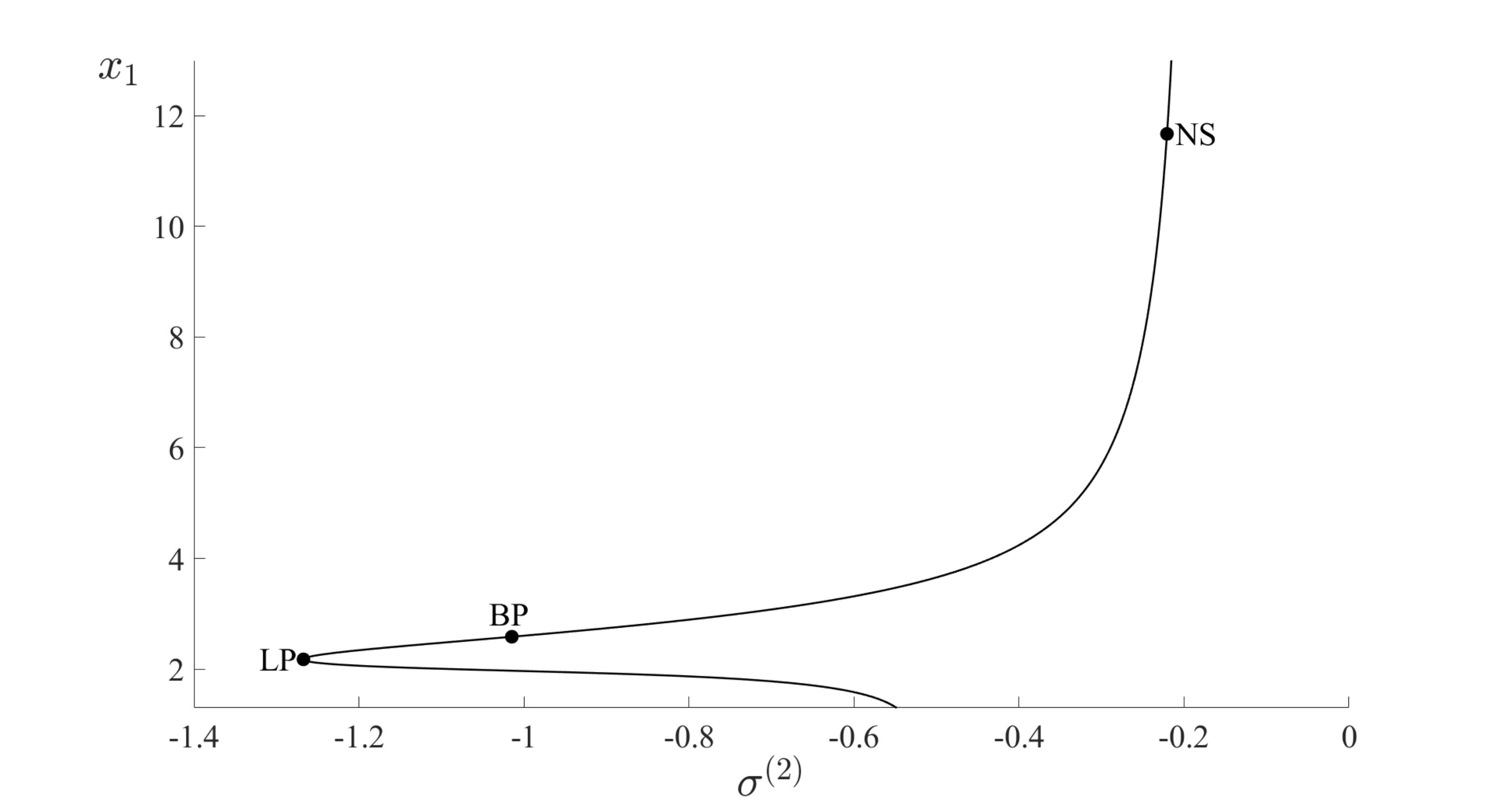}
 \caption{Bifurcation diagram of map \eqref{eq:model_p}-\eqref{eq:model_1} with $\sigma^{(2)}$ as bifurcation parameter with other parameters fixed. The black curve corresponds to the fixed points of the map. NS: Neirmark-Sacker, LP: limit point, and BP: branch point.}
 \label{fig:codim_sigma2}
\end{figure}

Apart from the period-doubling bifurcation, there are other codimension-1 bifurcations that can be observed in complex dynamical systems as our network model. Those cannot be detected directly with simulations using \texttt{Python}. To capture these bifurcations we use the numerical bifurcation tool for maps, \textsc{MatContM}~\cite{KuMe19}. We have implemented this to report codimension-$1$ bifurcation patterns like {saddle-node} bifurcation (sometimes called {\em fold} or {\em limit-point} bifurcation), {\em Neimark-Sacker} bifurcation, and {\em branch points} on the curves of the fixed point, in the negative $\sigma^{(2)}$ domain (see Fig.~\ref{fig:codim_sigma2}), which is otherwise not captured by \texttt{Python}. The coupling strengths are set as $\sigma^{(1)}=0.001$ and $\sigma^{(2)}=[-1.4,0.1]$, and the local parameters are set as $a=0.89, b=0.28, c=0.901, k_{0}=0.06,$ and $\mu=0.03$. Various codimension-$1$ and -$2$ patterns were also reported by Muni {\em et al.}~\cite{MuFa22} for a system of three-dimensional memristive Chialvo neuron. 

Codimension-$1$ bifurcations like period-doubling, saddle-node, and Neimark-Sacker occur when one of the eigenvalues of the Jacobian matrix $J$ has modulus $1$ at the fixed point $X^*$. Specifically, a saddle-node occurs when the eigenvalue is $1$, and a period-doubling occurs when the eigenvalue is $-1$ at $X^*$. When the eigenvalue is complex with modulus $1$, then a Neimark-Sacker bifurcation takes place. A {\em branch point} of fixed points is similar to a pitchfork bifurcation of equilibria in continuous systems. It appears in systems with symmetry or equivariance under variation of a single parameter \cite{Go00}.

A bifurcation diagram of the map \eqref{eq:model_p}-\eqref{eq:model_1} is shown in Fig.~\ref{fig:codim_sigma2}. We choose $\sigma^{(2)}$ as the bifurcation parameter because of its biological interest in understanding the dynamical behavior of a network of neurons. Varying $\sigma^{(2)}$ in the range $[-0.4, 0.1]$, three bifurcation points are detected -- NS: Neirmark-Sacker, LP: saddle-node (limit point), and BP: branch point. The supercritical Neirmark-Sacker with normal form coefficient of $-2.6556 e^{-01}$ occurs at $(x_{1},\sigma^{(2)})=(11.6726,-0.22066)$, the saddle-node bifurcation occurs at $(x_{1},\sigma^{(2)})=(2.1761,-1.2672)$ with normal form coefficient $1.2894e^{-02}$, and the branch point at $(x_{1},\sigma^{(2)})=(2.5847,-1.0147).$

\section{Time series analysis}
\label{sec:tsa}
Our network simulation will generate a battery of time-series data which is very rich in terms of complexity. Thus it becomes imperative to study this richness via crucial tools capturing the intricate trends. One such pattern is the onset of {\em chaos} from regular periodic solutions. We need tools to quantify these temporal behavior and two of these tools that we have implemented here are the $0-1$ test for chaos and {\em sample enrtopy} for complexity. We discuss these step by step in the next part of this section.

First, we describe the $0-1$ test in detail following Gottwald {\em et al.}~\cite{GoMe09}. We discuss the steps of the algorithm and how it has been implemented in our system. For time series analysis, we need time series data $\{x(n), n = 1, \ldots, \mathcal{N} \}$. To start with the 0-1 test, we need to first compute two translation variables $p_e$ and $q_e$ for a small value $e \in (0, 2\pi)$, given by
\begin{align}
    \label{eq:pe}
    p_e(n) = \sum_{i=1}^n x(i) \cos (ei), \\
    \label{eq:qe}
    q_e(n) = \sum_{i=1}^n x(i) \sin(ei),
\end{align}
for $n=1, \ldots, \mathcal{N}$. These can also be written as a two-dimensional discrete-time dynamical system as
\begin{align}
    \label{eq:pe2}
    p_e(n+1) = p_e(n) + x(n) \cos (en), \\
    \label{eq:qe2}
    q_e(n+1) = q_e(n) + x(n) \sin (en).
\end{align}
The phase portrait of~\eqref{eq:pe2}-\eqref{eq:qe2} will be typically bounded for regular dynamics (fixed point, periodic solution, and quasi-periodic orbit), whereas will be a two-dimensional approximation of the diffusive Brownian motion evolving with growth rate $\sqrt{n}$ (zero drift). To mathematically diagnose whether $p_e$ and $q_e$ behave diffusively or not, we employ the mean square displacement $M_e$ given by
\begin{align}
    M_e(n) = \frac{\sum_{i=1}^\mathcal{N} \left\{p_e(i+n) - p_e(i) \right\}^2 + \left\{q_e(i+n) - q_e(i) \right\}^2}{\mathcal{N}}.
\end{align}
If $M_e$ grows linearly, then the system~\eqref{eq:pe2}-\eqref{eq:qe2} is diffusive, otherwise is bounded if $M_e$ is bounded. For the purpose of the numerics, it is required that $n \ll \mathcal{N}$. Thus, $M_e(n)$ is evaluated only for an $n \le \mathcal{N}_{\rm crit}$, where $\mathcal{N}_{\rm crit}$ is a critical value with $\mathcal{N}_{\rm crit} \ll \mathcal{N}$. It is recommended to keep $\mathcal{N}_{\rm crit} \le \frac{\mathcal{N}}{10}$. Next, the mean square displacement was modified by adding a correction term $M_{\rm correc, e}$~\cite{GoMe09b} to $M_e$. i.e,
\begin{align}
\label{eq:De}
D_e(n) = M_e(n) +M_{{\rm correc}, e}(n),
\end{align}
where
\begin{align}
    M_{{\rm correc}, e}(n) = -\left(\lim_{\mathcal{N} \to \infty} \frac{1}{\mathcal{N}} \sum_{i=1}^\mathcal{N} x(i) \right)^2 \frac{1 - \cos(en)}{1 - \cos (e)}.
\end{align}
This modified version $D_e$ not only has better convergence properties but also reveals the same asymptomatic growth rate as $M_e$. The next step in this algorithm is to evaluate this asymptomatic growth rate given by
\begin{align}
    K_e = \lim_{n \to \infty} \frac{\log M_e(n)}{\log n}.
\end{align}
As put forward by Gottwald {\em et al.}~\cite{GoMe09}, there are two possibilities to compute this $K_e$: by {\em regression} and by {\em correlation}. We follow the first approach for this paper. The regression method is applied to $D_e$ rather than $M_e$ as $D_e$ portrays lesser variance than $M_e$, see Fig.~\ref{fig:MDK}.  Because $D_e$ might be negative, it is imperative to modify $D_e$ as
\begin{align}
\label{eq:De_tilde}
\tilde{D}_e(n) = D_e(n) - \min_{1\le n \le \mathcal{N}_{\rm crit}} |D_e(n)|.
\end{align}
Then $K_e$ will be numerically evaluated by regression method from the graph of $\log \tilde{D}_e(n)$ vs. $\log n$. If $M_e$ grows linearly, $K_e \approx 1$, indicating chaotic dynamics in the system, whereas if $M_e$ is bounded, $K_e \approx 0$, indicating regular behavior. Finally several of these $K_e$'s should be computed for various $e$ values. From these values, a statistically more accurate $K$ value is computed, given by $K = {\rm median} (K_e)$.

\begin{figure}[h]
\centering
\begin{tabular}{cc}
  \includegraphics[scale=0.2]{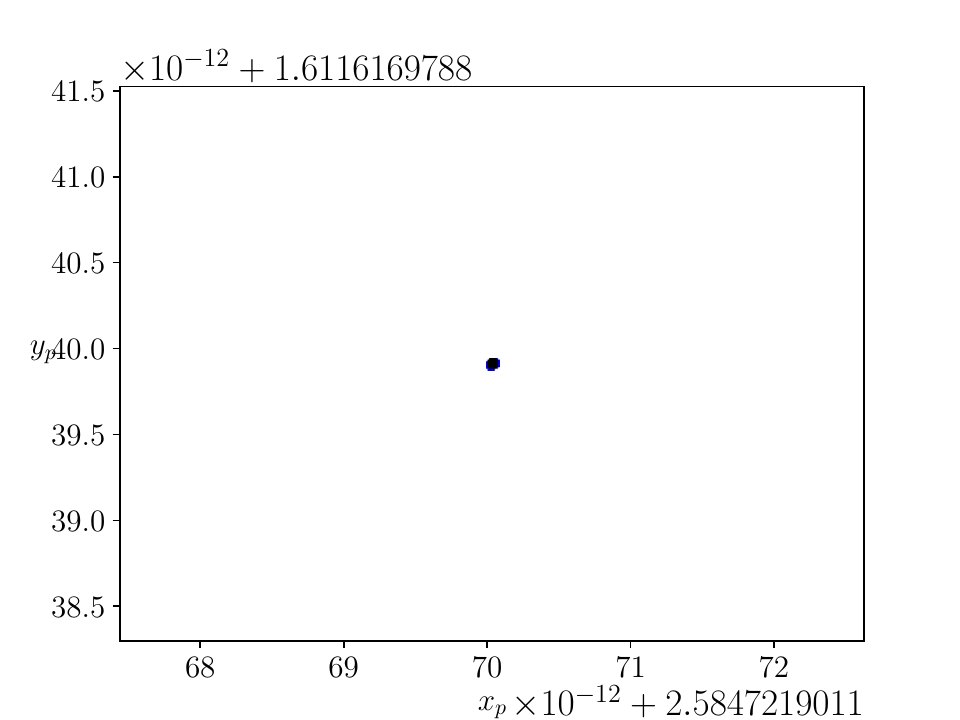} &   \includegraphics[scale=0.2]{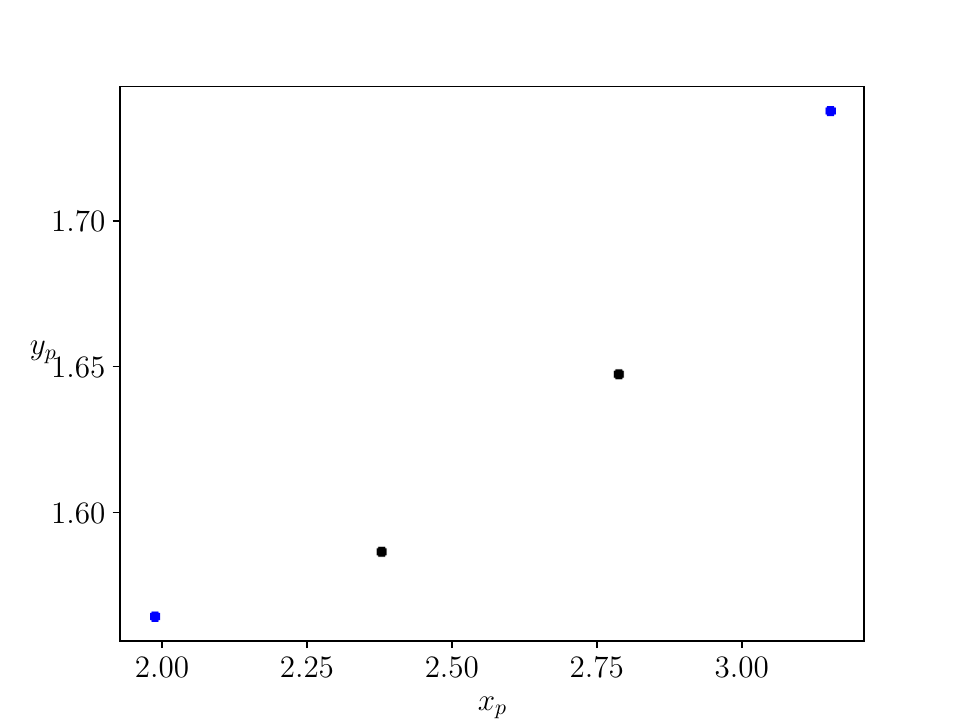} \\
(a) $\sigma^{(2)} = 0.08$ & (b) $\sigma^{(2)} = 0.09$ \\[3pt]
\includegraphics[scale=0.2]{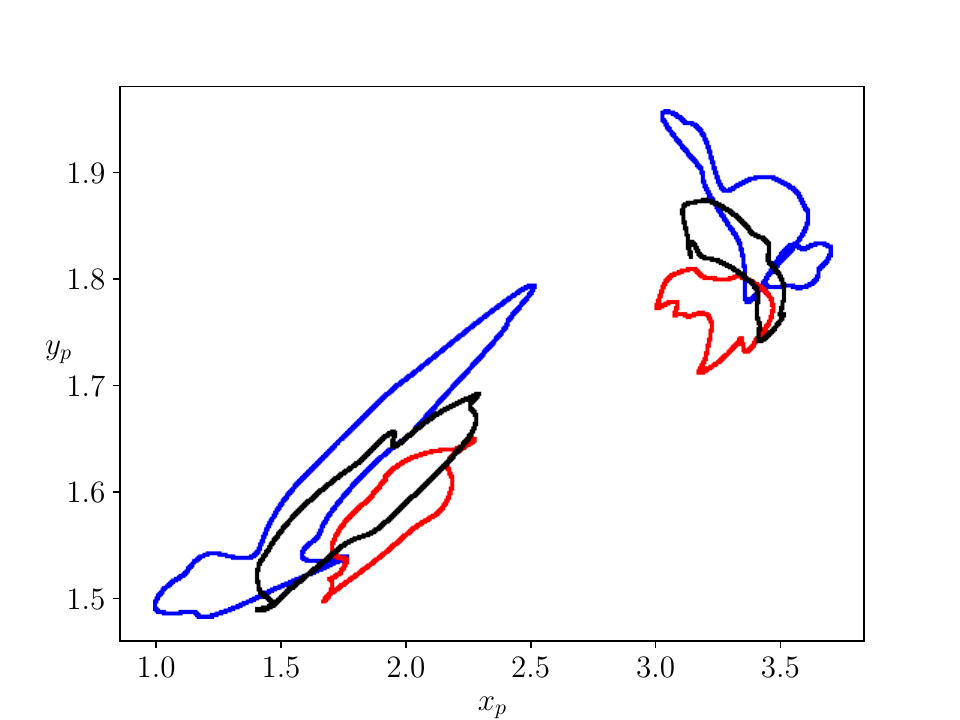} &   \includegraphics[scale=0.2]{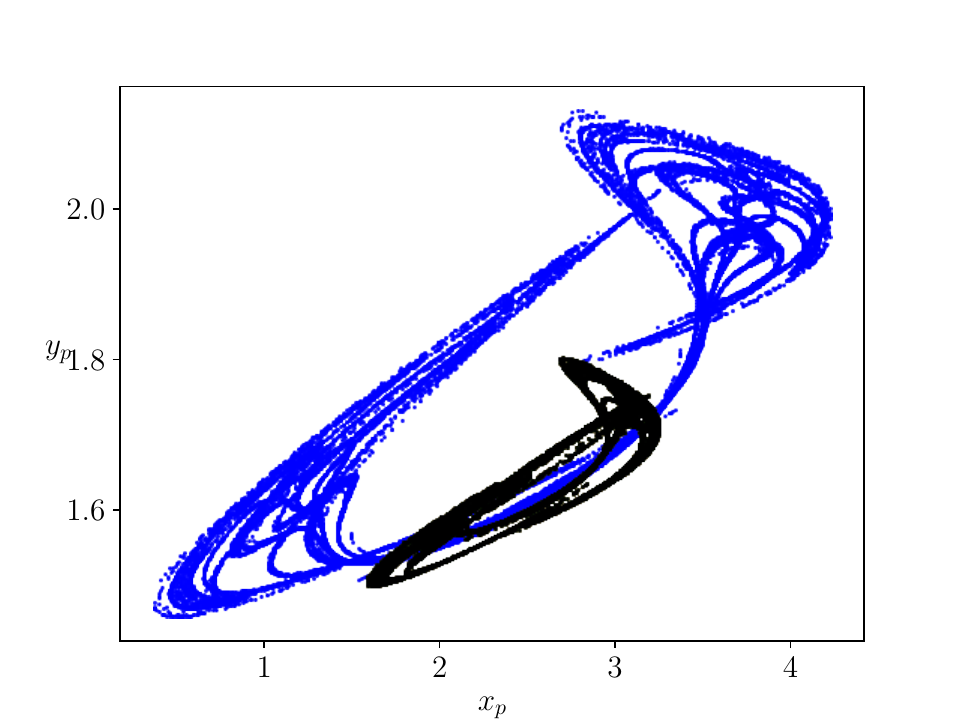} \\
(c) $\sigma^{(2)} = 0.11$ & (d) $\sigma^{(2)} = 0.115$ \\[3pt]
\end{tabular}
\caption{Typical phase portraits of the system~\eqref{eq:model_p}-\eqref{eq:model_1}. The simulations are run for $50000$ iterations, out of which the last $30000$ are plotted to ensure transients are discarded. Parameter values are fixed as $a=0.89$, $b=0.28$, $c=0.901$, $k_0 = 0.06$, $\mu=0.03$, and $\sigma^{(1)}=0.001$, with $\sigma^{(2)}$ as the primary bifurcation parameter. Initial conditions are randomly sampled from the uniform distribution $[0.6, 0.8]$. Panel (a) represents the phase portrait for $\sigma^{(2)}=0.08$ and shows a fixed point. Panel (b) is for $\sigma^{(2)}=0.09$, showing period-doubling. Panel (c) for $\sigma^{(2)}=0.11$, exhibits a disjoint cyclic quasiperiodic closed invariant curve. Finally, (d) represents $\sigma^{(2)}=0.115$, exhibiting chaos.}
\label{fig:bif_pp}
\end{figure}

\begin{figure}[h]
\centering
\begin{tabular}{c}
  \includegraphics[scale=0.18]{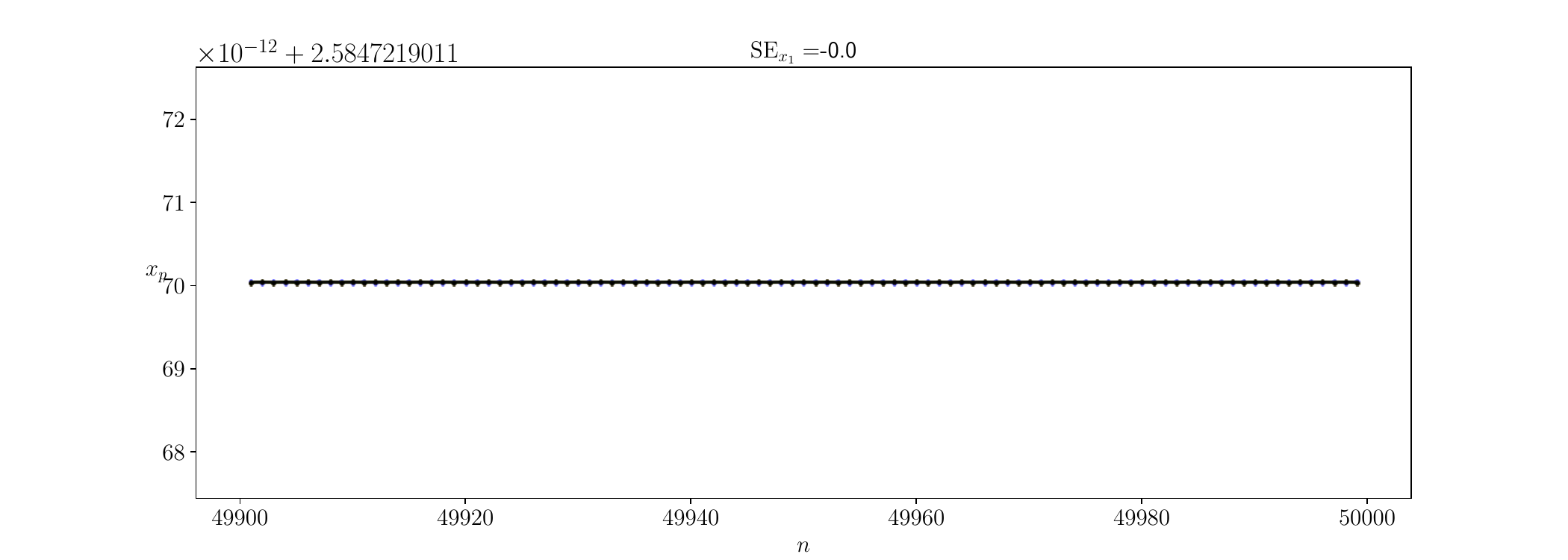} \\
  (a) $\sigma^{(2)} = 0.08$ \\
  \includegraphics[scale=0.18]{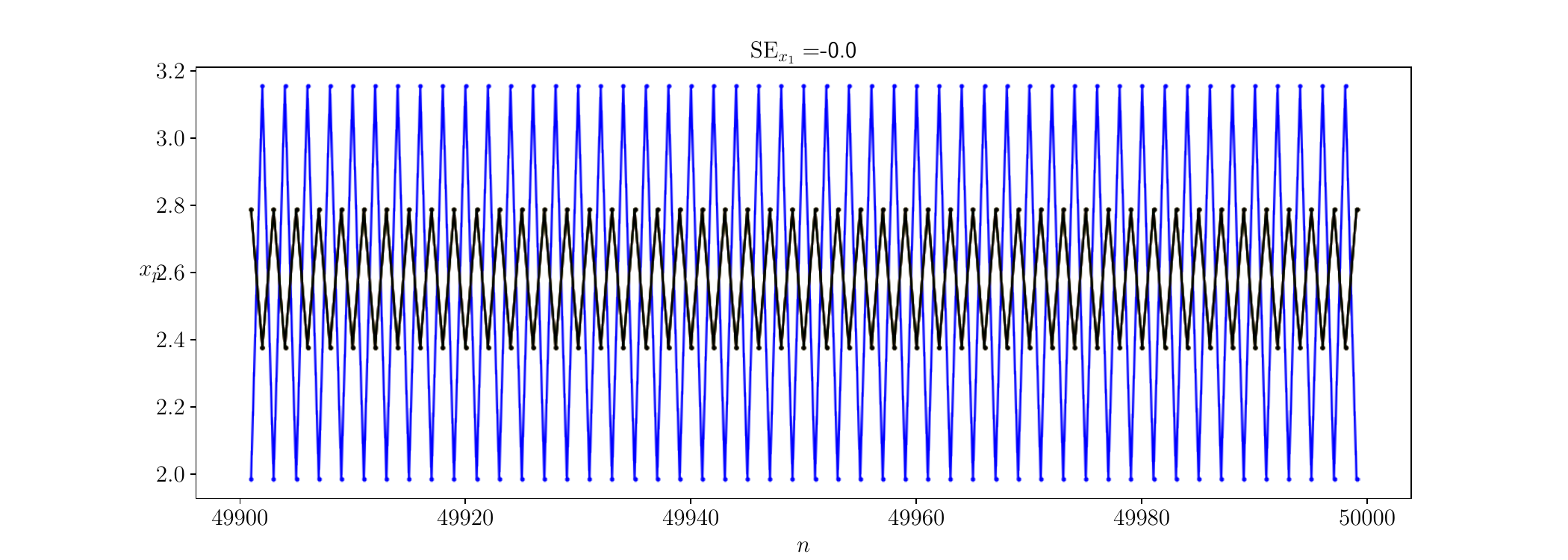}\\[3pt]
  (b) $\sigma^{(2)} = 0.09$ \\
  \includegraphics[scale=0.18]{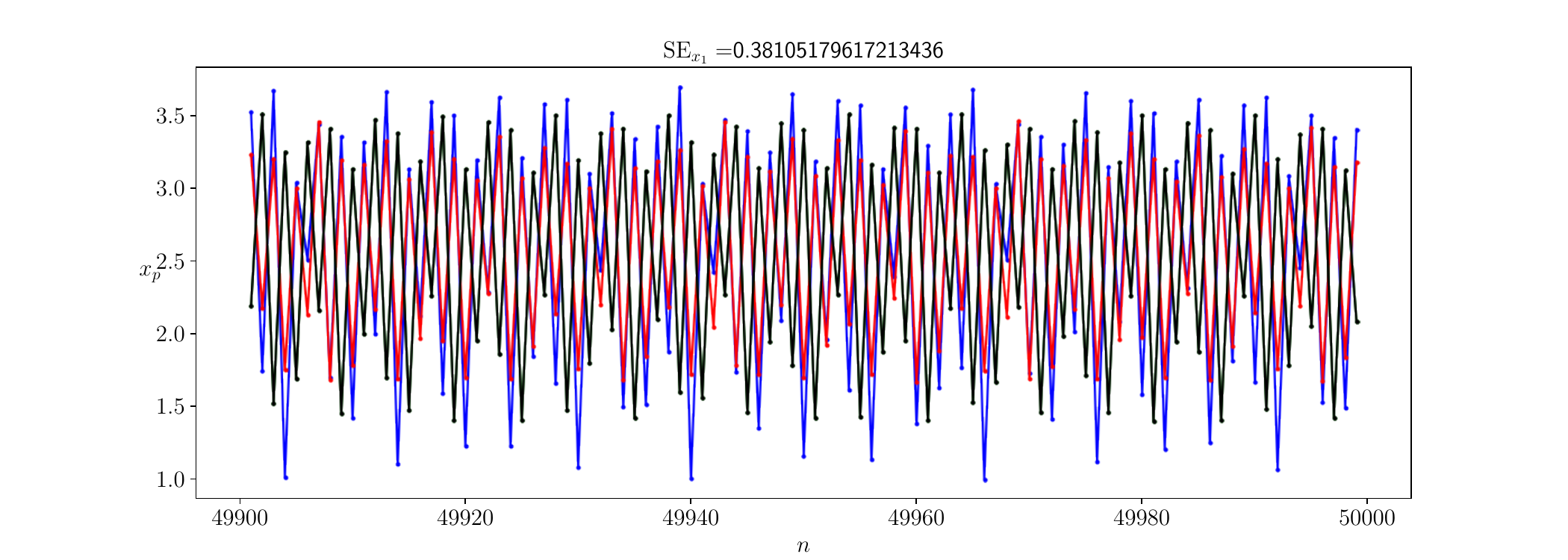} \\
  (c) $\sigma^{(2)} = 0.11$ \\
  \includegraphics[scale=0.18]{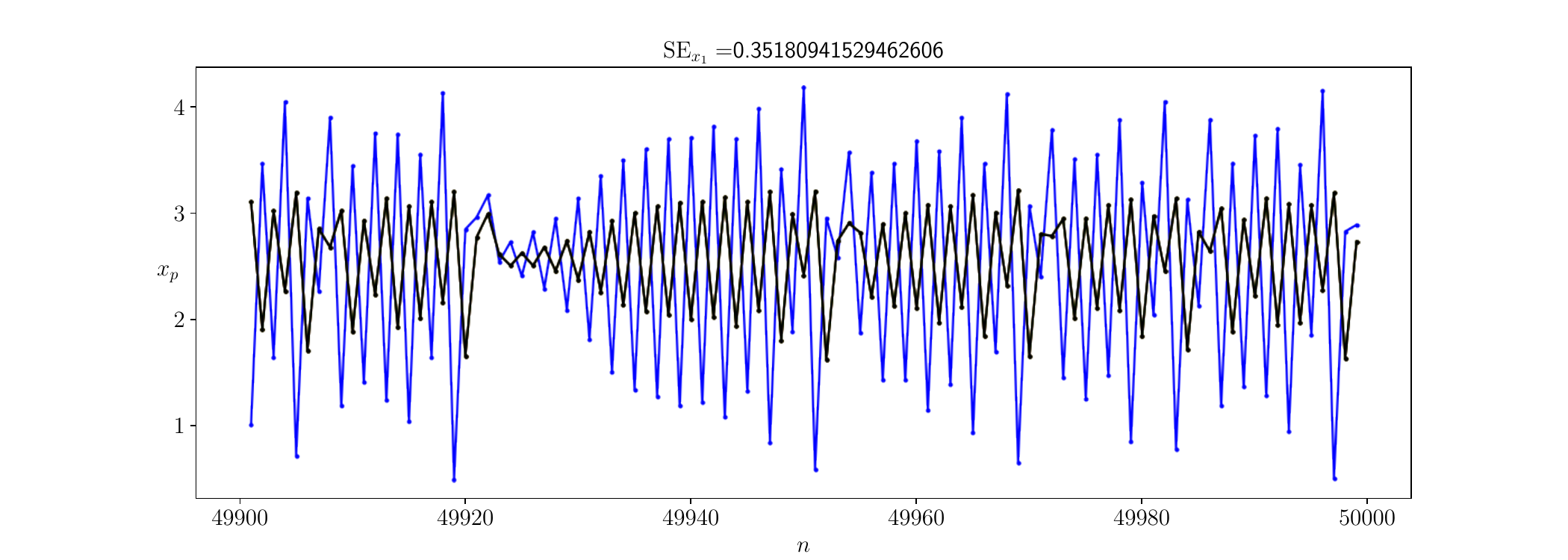}\\[3pt]
  (d) $\sigma^{(2)} = 0.115$
\end{tabular}
\caption{Time series plots corresponding to the phase portraits in Fig.~\ref{fig:bif_pp}. Panel (a) is for fixed point, (b) for period-doubling, (c) for transition to chaos, and (d) for chaos.}
\label{fig:bif_TS}
\end{figure}

\begin{figure}[h]
\centering
\begin{tabular}{cc}
  \includegraphics[scale=0.2]{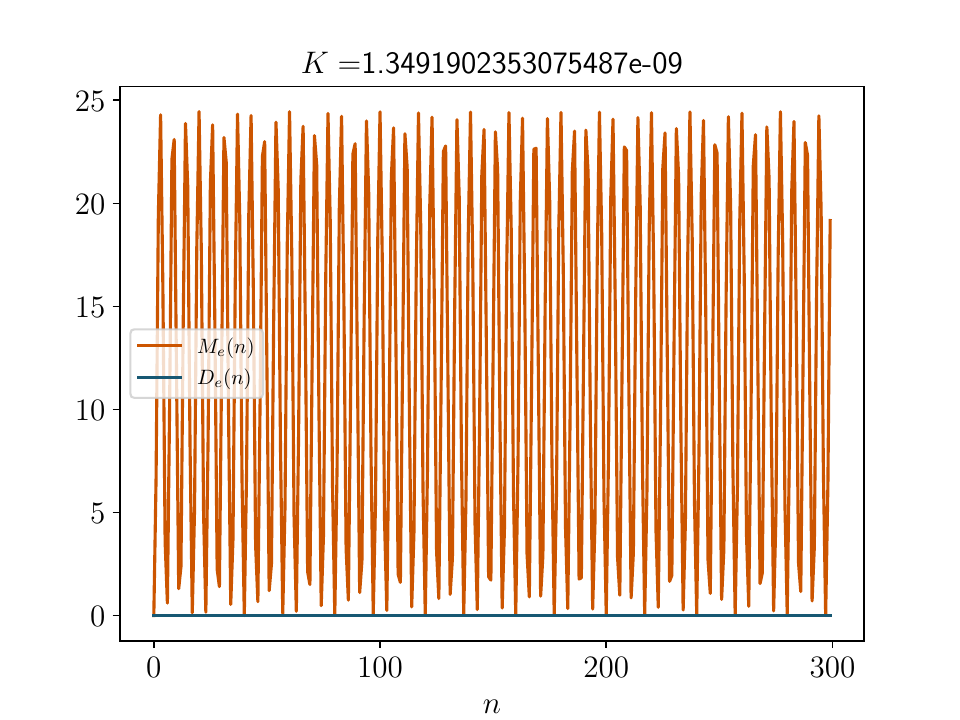} &   \includegraphics[scale=0.2]{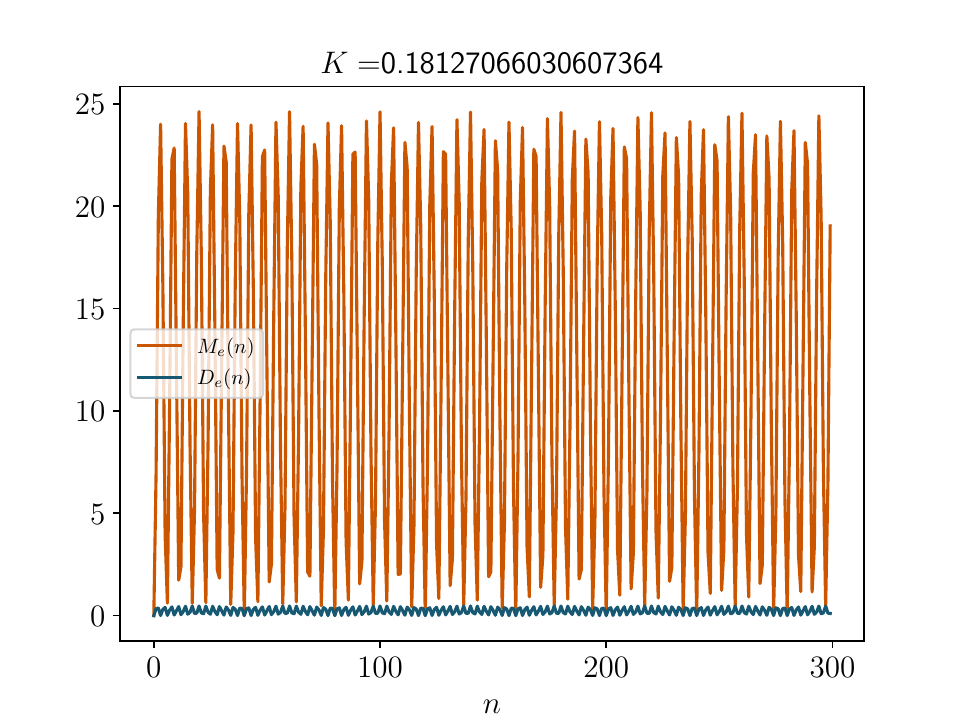} \\
(a) $\sigma^{(2)} = 0.08$ & (b) $\sigma^{(2)} = 0.09$ \\[3pt]
\includegraphics[scale=0.2]{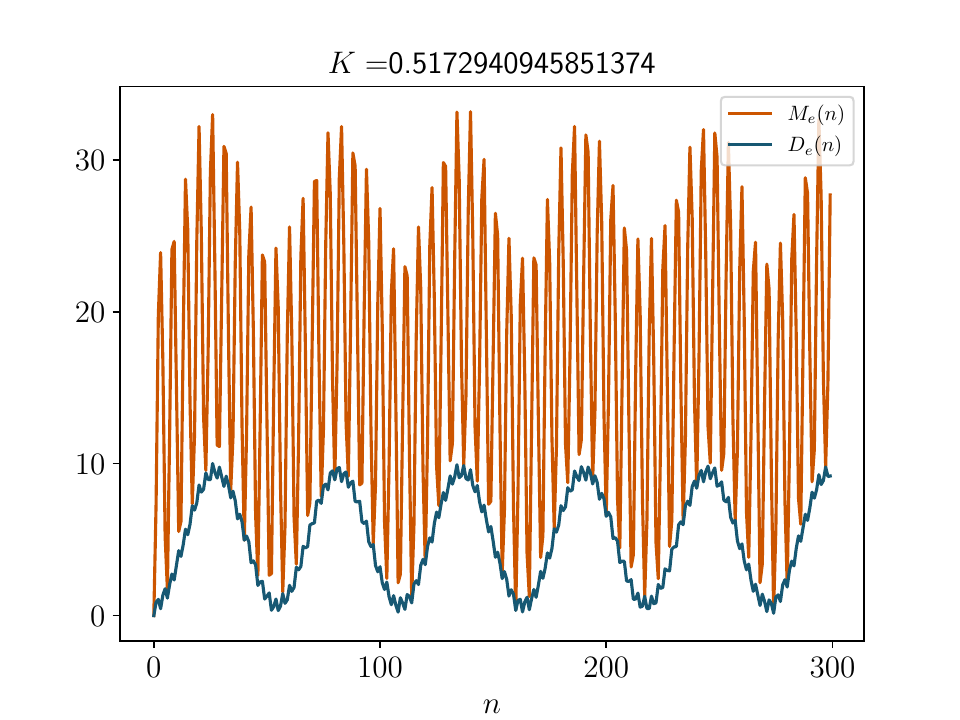} &   \includegraphics[scale=0.2]{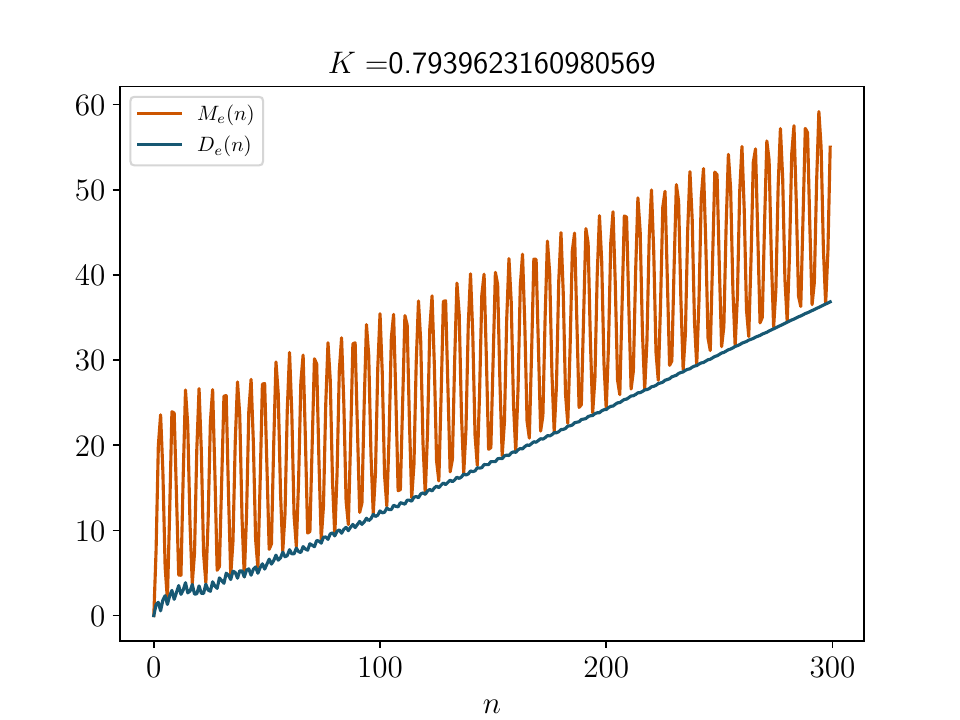} \\
(c) $\sigma^{(2)} = 0.11$ & (d) $\sigma^{(2)} = 0.115$ \\[3pt]
\end{tabular}
\caption{Plot of $M_c$ and $D_c$ at different $\sigma^{(2)}$ values corresponding to the phase portraits in Fig.~\ref{fig:bif_pp}, with $K$ values noted. Panel (a) is for fixed point with $K \approx 0$, (b) for period-doubling with $K \approx 0.181$, (c) for the chaotic gap with $K \approx 0.517$, and (d) for chaos, with $K\approx 0.794$.}
\label{fig:MDK}
\end{figure}

First, we consider four parameter values of interest from the bifurcation diagram Fig.~\ref{fig:bif}, given by $\sigma^{(2)} = 0.08, 0.09, 0.11, 0.115$ (indicated by the blue dashed lines). At first look, intuition tells us that at $\sigma^{(2)}=0.08$ we have a fixed point, at $\sigma^{(2)}=0.09$, we have a period-doubling, and at $\sigma^{(2)}=0.11$ and $0.115$, we have chaos. The typical phase portraits for these four scenarios are given in Fig.~\ref{fig:bif_pp}. To generate these phase portraits, we have followed the same simulation steps as that of Fig.~\ref{fig:bif}. Reiterating, the simulation is run for $5 \times 10^4$ iterations and only the last $3 \times 10^4$ points are plotted as $x_p$ vs. $y_p$, to ensure transients are discarded. Note that following the color convention in our schematic Fig.~\ref{fig:schematic}, we have colored the nodes in the phase portraits, i.e., $1$: blue, $2$: red, $3$: green, and $4$: black. The initial values of the dynamical variables $x_1 \to y_4$ are sampled uniformly from the range $[0.6, 0.8]$. Looking at the phase portraits, we observe that panel (a) is indeed a fixed point, and panel (b) is a period-doubling. Panel (c) shows the loss of smoothness of a disjoint cyclic quasiperiodic closed invariant curve. Such loss of smoothness usually shows the beginning of the transition towards the formation of a chaotic attractor. Further, an increase in $\sigma^{(2)}$ results in the two disjoint cyclic components approaching each other and merging into a single chaotic attractor via an attractor merging crisis at $\sigma^{(2)} = 0.115$, see panel (d). We notice four dots two colored in blue and the other two in black. This readily corroborates with the time series plots~\ref{fig:bif_TS}-(a) and (b). We notice that the central node (node $1$) oscillates with an amplitude different from the three peripheral nodes which are in complete synchrony on their own. Fig.~\ref{fig:bif_TS}-(a) is a straight line (fixed point) and Fig.~\ref{fig:bif_TS}-(b) shows regular patterns in two colors indicating node $1$ oscillates in a different amplitude to the other three. Now, the third phase portrait exhibits an interesting phenomenon, where the attractors have split into two pieces. This behavior is anticipated as a transition toward chaos. Nodes $1$, $2$, and $4$ oscillate in different amplitudes from each other, however, the trajectory of node $3$ is in tandem with node $4$. This is also clearly indicated in the fairly irregular time series bursts in Fig.~\ref{fig:bif_TS}-(c). Finally, Fig.~\ref{fig:bif_pp}-(d) exhibits chaos, where the attractor for the central node is topologically distinguished from the attractors for the other nodes (node $2$, and $3$ are masked by node $4$). The peripheral nodes are in complete synchrony with each other. The time series~\ref{fig:bif_TS}-(d) is highly irregular indicating an onset of chaos. It should be noted that only the last $100$ iterations are illustrated for Fig.~\ref{fig:bif_TS} to make the plots comprehensible. Our next task is to confirm all these regular and chaotic behaviors numerically via the $0-1$ test.

\begin{figure}[h]
\centering
\begin{tabular}{cc}
  \includegraphics[scale=0.2]{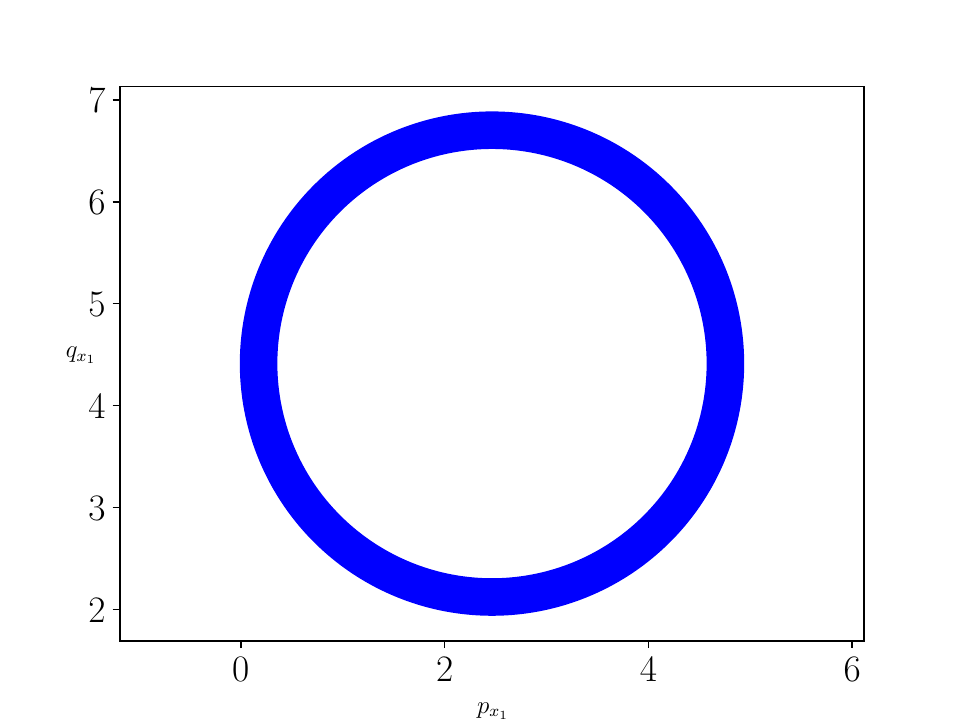} &   \includegraphics[scale=0.2]{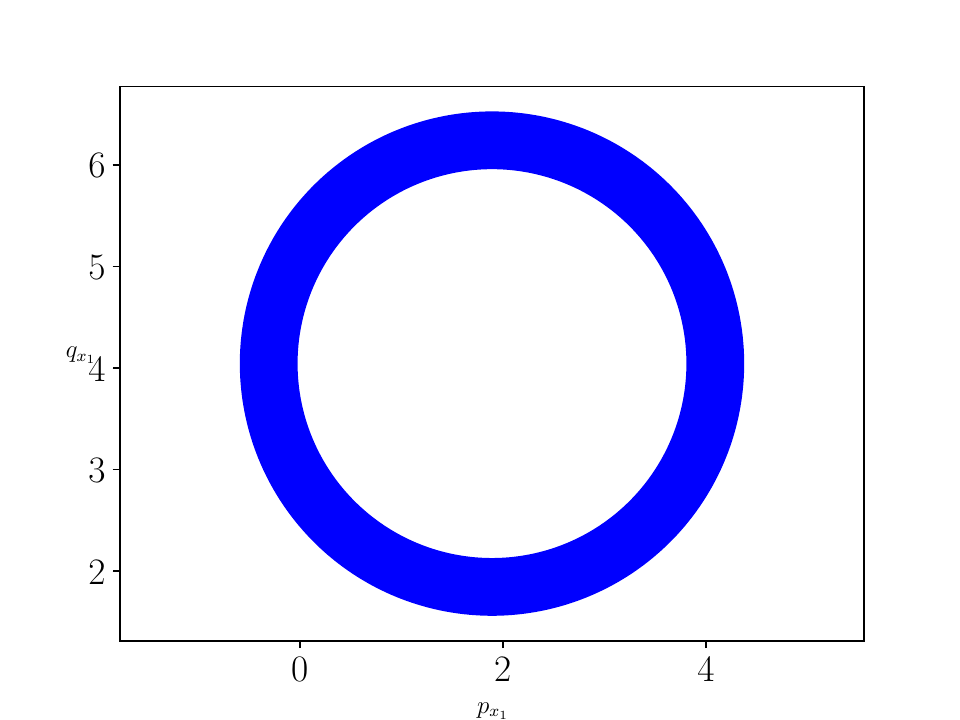} \\
(a) $\sigma^{(2)} = 0.08$ & (b) $\sigma^{(2)} = 0.09$ \\[3pt]
\includegraphics[scale=0.2]{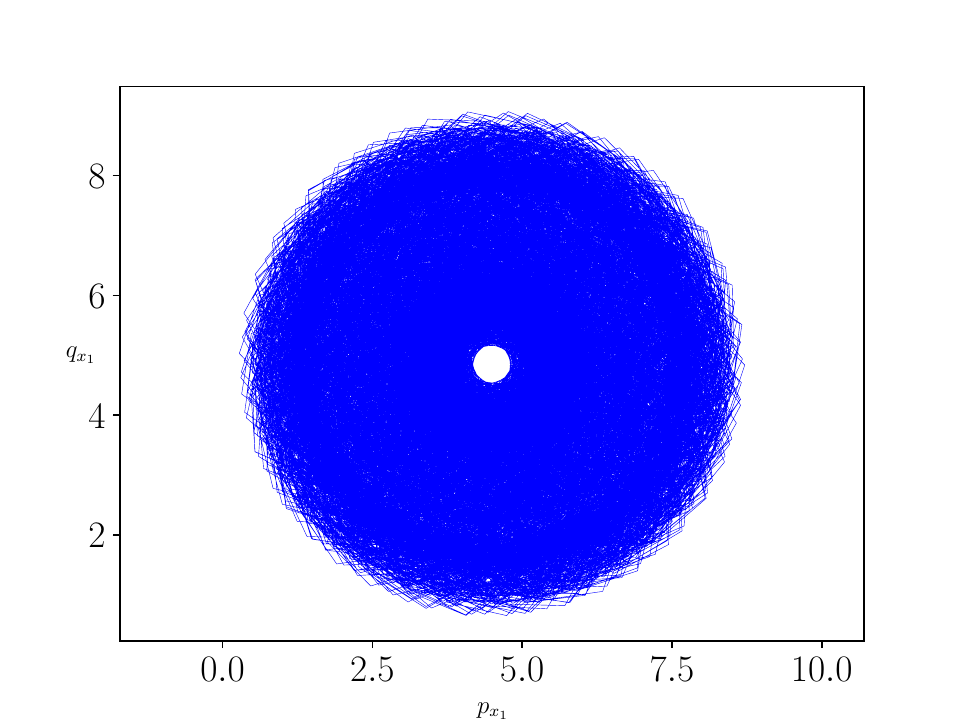} &   \includegraphics[scale=0.2]{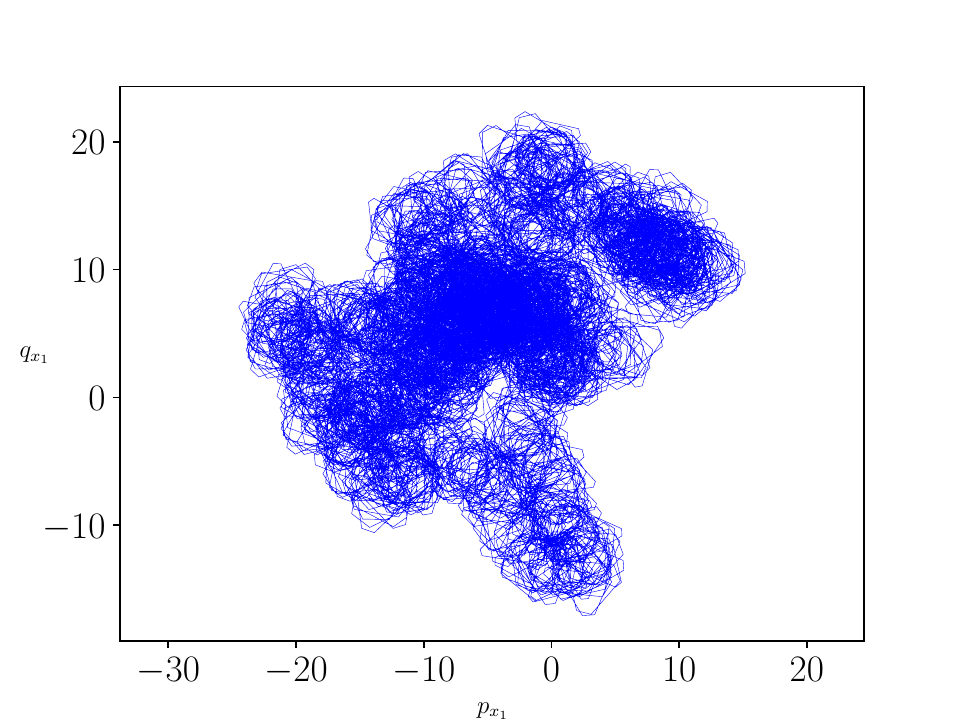} \\
(c) $\sigma^{(2)} = 0.11$ & (d) $\sigma^{(2)} = 0.115$ \\[3pt]
\end{tabular}
\caption{Plot of $p_{x_1}$ versus $q_{x_1}$ at different $\sigma^{(2)}$ values corresponding to the phase portraits in Fig.~\ref{fig:bif_pp}. Panel (a) shows a highly bounded trajectory, whereas (b) shows a slightly less bounded trajectory. In (c) we see a trajectory that lies between bounded and diffusive, and in (d) we see a diffusive random walk corresponding to a Brownian motion with zero drift.}
\label{fig:pq}
\end{figure}

The first step is to compute the translation variables $p_e$ and $q_e$. For the above four cases, we have chosen $e=1.1$. As noted before, the simulations are run for $5 \times 10^4$ iterations, out of which the first $2 \times 10^4$ are discarded. This ensures that the time series data does not contain any transients. Thus we have $\mathcal{N} = 3 \times 10^4$. Then we plot $p_{x_1}$ versus $q_{x_1}$ showing the phase portraits of~\eqref{eq:pe2} and~\eqref{eq:qe2} for the central node (node 1), see Fig.~\ref{fig:pq}. The trajectories are colored blue following the color code of the nodes set in Fig.~\ref{fig:schematic}. We only show these for the central node because the qualitative behavior of the other nodes will follow a similar pattern at the set parameter values. Thus focusing on only one node suffices. We observe that the trajectory is highly bounded for the fixed point (panel (a), $\sigma^{(2)}=0.08$), whereas it is slightly less bounded for the period-doubling case (panel (b), $\sigma^{(2)}=0.09$). When it comes to the behavior at $\sigma^{(2)}=0.11$ exhibiting transition to chaos, the trajectory lies somewhat in between bounded and diffusive. This also qualitatively indicates a transition to chaos. Finally, at $\sigma^{(2)}=0.115$, where chaos is observed, the trajectory is a random walk showing a diffusive nature similar to a Brownian motion with zero drift. This brings us to mathematically analyzing the diffusive patterns of $p_e$ and $q_e$ via the computation of $M_e$ and the modified version $D_e$. To do that we need to set $\mathcal{N}_{\rm crit} \le \frac{\mathcal{N}}{10}=3 \times 10^3$. Let us consider two values for this term, i.e $\mathcal{N}_{\rm crit}=300$, and $\mathcal{N}_{\rm crit}=50$. We use the first one to plot the graphs for $M_e$ and $D_e$, and the second one to ultimately compute $K$. Note that a small value of $\mathcal{N}_{\rm crit}$, makes the graphs of $M_e$ and $D_e$ look blocky, but makes it faster to compute $K$, whereas, a large value of $\mathcal{N}_{\rm crit}$ makes the graphs of $M_e$ and $D_e$ look less blocky but makes the computation of $K$ slower. These caveats require us to consider two values. We plot $M_e$ and $D_e$ and also compute $K$ showing the results in Fig.~\ref{fig:MDK}. Note that in panel (a) where $\sigma^{(2)}=0.08$ (fixed point), both $M_e$ and $D_e$ are highly bounded as expected, with $K=0$. For $\sigma^{(2)}=0.09$, we see that in panel (b), the $M_e$ and $D_e$ plots are slightly less bounded as compared to the ones in panel (a) (although difficult to say with bare eyes). In this case, we have a slightly higher value of $K \approx 0.181$ pertaining to period doubling, which is regular by itself, but is less regular when compared geometrically to a fixed point. For panel (c) where we are observing a transition to chaos (at $\sigma^{(2)} = 0.11$), we start seeing subtle changes in the growth rates of $M_e$ and $D_e$ which are no longer highly bounded. The $K$ value increases to approximately $0.517$, indicating a transition towards chaotic behavior. Finally for $\sigma^{(2)} = 0.115$, the growth rates of $M_e$ and $D_e$ become linear indicating chaos, with $K \approx 0.794$. The simulations for the $0-1$ test have been achieved by converting the \texttt{Julia} code \texttt{01ChaosTest.jl}~\cite{amJulia} to \texttt{Python}. 

After the test of chaoticity, it is time to look into a test for complexity in the time series. We implement this via {\em sample entropy} following {\em Richman et al.}~\cite{RiMo00}. This gives us a quantifier to measure the rate of information production in the network system over time. First, we detail the algorithm for the sample entropy in terms of a general time series before implementing it in our model. Given the time series $\{x(n), n=1, \ldots, \mathcal{N} \}$, we first define a vector $x_v(j)$ as
\begin{align}
\label{eq:xvj}
x_v(j) = \left\{x(j+k) \mid 0 \le k \le v-1 \right\},\; 1 \le j \le \mathcal{N}-v+1,
\end{align}
for a non-negative integer $v \le \mathcal{N}$. Note that there exist $\mathcal{N} -v+1$ of these sets and each of these consists of $v$ data points ranging from $x(j)$ to $x(j+v - 1)$. The next step is to compute the Euclidean distance given by
\begin{align}
\mathcal{E} \left(x_v(j), x_v(n)\right) = \max_{0 \leq k \leq v-1} \left\{\lvert x(j+k) - x(n+k) \rvert \right\}. \nonumber
\end{align}
Thereafter, we require a real-valued threshold $\delta >0$ utilising which we calculate the term $B^v_j(\delta)$ given by the ratio of the number of vectors $x_v(j)$ satisfying the inequality $\mathcal{E} \left(x_v(j), x_v(n)\right) \le \delta$ to the number $\mathcal{N}-v-1$. We also have $1 \le n \le \mathcal{N}-v$ with $n \ne j$. Thus the mean is computed as
\begin{align}
B^v(\delta) = \frac{1}{\mathcal{N}-v} \sum_{j=1}^{\mathcal{N} - v} B_j^v(\delta). \nonumber
\end{align}
Using the same technique we can also compute $B^{v+1}(\delta)$, ultimately giving us the sample entropy measure of the time series as
\begin{align}
\label{eq:sampen}
{\rm SE} = \lim_{\mathcal{N} \to \infty}\left(-\ln \frac{B^{v+1}(\delta)}{B^v(\delta)} \right).
\end{align}
A low value of ${\rm SE}$ is indicative of lower unpredictability in the system pointing towards a low complexity, whereas a high value indicates higher complexity in the system. Intuition says that the higher the chaoticity of a dynamical system, the higher its complexity will be. We implement sample entropy to various time series from our model utilizing the open-source \texttt{Python} package called \texttt{nolds}~\cite{Sc19}. This has also been recently used to evaluate the sample entropy of a heterogeneous ring-star network of memristive Chialvo neurons in the thermodynamic limit by Ghosh {\em et al.}~\cite{GhMu23}. Like in the $0-1$ test, we simulate the model for $5\times 10^4$ time steps. But this time to evaluate sample entropy, we discard the first $2.5 \times 10^4$, making $\mathcal{N} = 2.5 \times 10^4$. The package \texttt{nolds} provides us a function \texttt{sampen()} that has been written specifically to perform the task of sample entropy from a time series data following the algorithm by Richman {\em et al.} The default value of $v$ and $\delta$ in this function has been set to be $2$ and $0.2$ times the standard deviation of the time series data. The corresponding sample entropy values for the central node of our network for four different $\sigma^{(2)}$ have been reported in Fig.~\ref{fig:bif_TS}. The reasoning behind considering only the central node lies parallel to the one discussed in the $0-1$ test. We see that when $\sigma^{(2)}=0.08$, the attractor is a fixed point and as expected the system has very regular dynamics, and thus very low complexity, given by ${\rm SE}=0$. This also occurs for the regular dynamics of period doubling, when $\sigma^{(2)}=0.09$, with ${\rm SE}=0$. On the further increase of $\sigma^{(2)}$ to $0.11$, we notice the transition to chaotic behavior supported by the increase in ${\rm SE}$ value to approximately $0.381$, i.e., high complexity. Finally for $\sigma^{(2)} = 0.115$ where the model exhibits the onset of chaos, very high complexity is again portrayed, supported by ${\rm SE} \approx 0.352$.

\begin{figure}[b!]
\vskip 6pt
\centering
\includegraphics[width=0.6\linewidth]{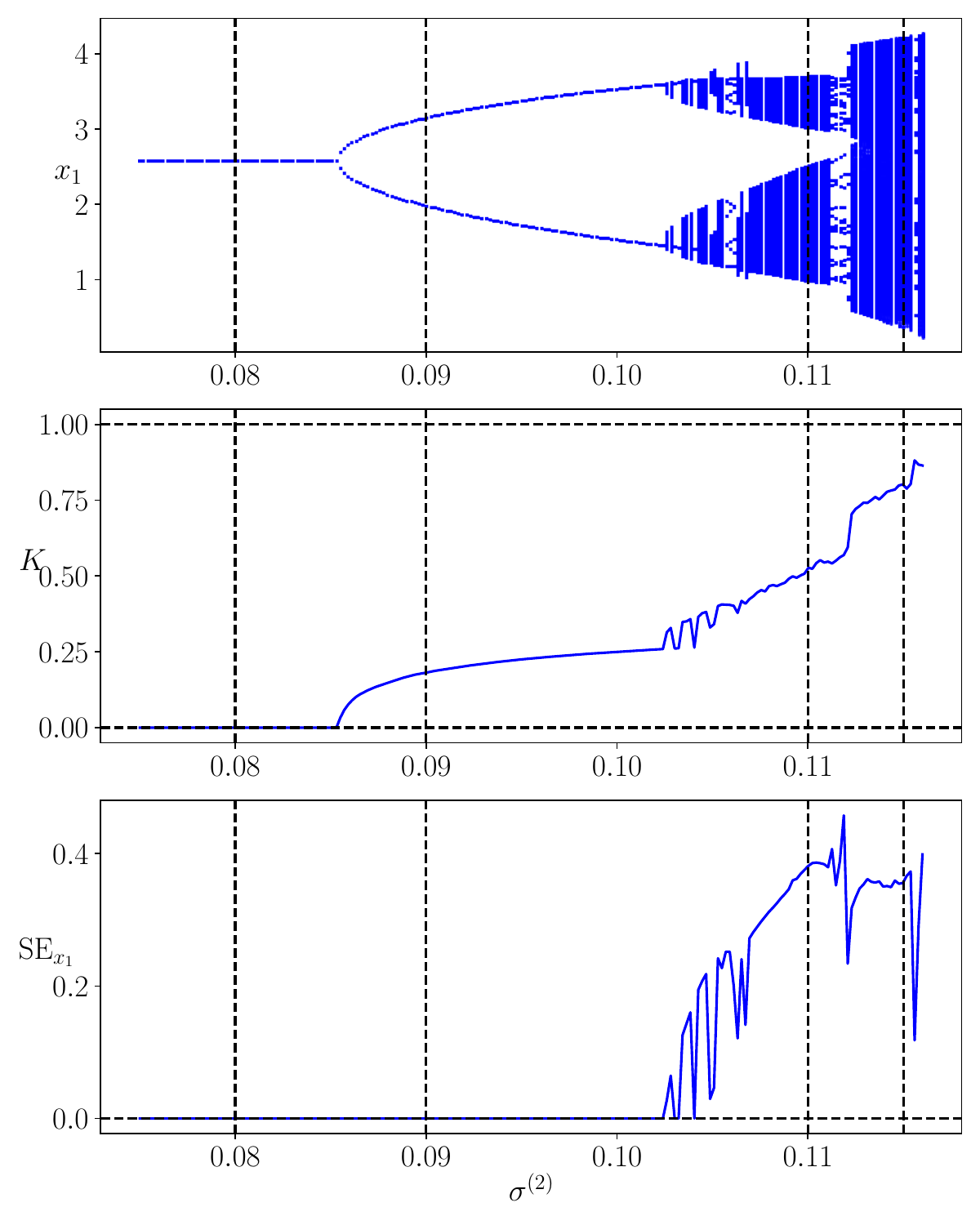}
\caption{A bifurcation plot of the first node with $\sigma^{(2)}$ as the main bifurcation parameter. The corresponding value of $K$ from the chaos test and ${\rm SE}$ for complexity are shown. The parameter values are set as $a=0.89$, $b=0.28$, $c=0.901$, $k_0=0.06$, $\mu=0.03$, and $\sigma^{(1)}=0.001$. The initial values of all the dynamical variables are sampled uniformly from $[0.6, 0.8].$}~\label{fig:K_se_plot}. 
\end{figure}

The final bit in this section is to illustrate a bifurcation diagram of $K$ and ${\rm SE}$ over a wide range of $\sigma^{(2)}$ values giving us a full picture of how they vary from regular dynamics to chaotic dynamics, considering $\sigma^{(2)}$ as the main bifurcation parameter. This is shown in Fig.~\ref{fig:K_se_plot}. This also gives us a statistical viewpoint on the relationship between the chaoticity and the complexity of a dynamical system. We recreate the bifurcation diagram of the dynamical variable $x_1$ with varying $\sigma^{(2)}$ in the range $[0.075, 0.115]$ in the top row of Fig.~\ref{fig:K_se_plot}. The $K$ plot (in the middle row) shows that $K=0$ for the part where we notice a fixed point. As soon as there is an onset of period-doubling, we observe that $0<K<0.5$, still indicating regularity in the dynamics. On the further increase of $\sigma^{(2)}$, as soon as there is an irregularity in the dynamics of the network (realized via transition to chaos and ultimately chaos), the $K$ value starts increasing beyond $0.45$ and reaches approximately $0.8$. Now for the ${\rm SE}$ plot (in the third row), we see ${\rm SE} = 0$ for all regular dynamics (fixed point, period-doubling). As soon as some irregularity sneaks into the dynamics, the ${\rm SE}$ value starts increasing exponentially, indicating higher complexity with higher irregularity. The vertical black dashed lines refer to the $\sigma^{(2)}$ values of interest at which we have plotted the phase portraits, time series, $p_{x_1}$ vs. $q_{x_1}$ portraits, and the $M_e$ and $D_e$ curves before. We can see via numerical simulations that chaos and complexity are intertwined and provide us a benchmark for further research on their unification, and what kind of questions they might give rise to in the context of mathematical biology and healthcare~\cite{RiHa07}. It would be interesting to have further research done on the mathematical framework of relating these two ubiquitous concepts.

\section{Synchronization Measures}
\label{sec:sync}
In this section, we numerically unravel the extent of synchronization in our model. We achieve this via two metrics: the cross-correlation coefficient and the Kuramoto order parameter. The reasoning behind taking two metrics is to corroborate the numerical result from one with that from the other. This gives us an overview of the collective dynamics of the network model where each node acts as an agent, organized with each other through pairwise and higher-order coupling strengths. 

\subsection{Cross-correlation coefficient}
Keeping in mind that our model is a ring-star network with one node at the center and three nodes on the periphery making a complete graph, we have to devise the cross-correlation coefficient in such a way that the correct collective dynamics are captured. Intuitively, a collection of six pairs will arise in terms of the geometry of the network. These will give six cross-correlation coefficients. The cross-correlation coefficient between a node $i$ and a node $j$ is given by
\begin{align}
\label{eq:Gamma_ij}
    \Gamma_{ij} = \frac{\langle\tilde{x_i}(n)\tilde{x_j}(n)\rangle}{\sqrt{\langle\tilde{x_i}(n)^2\rangle\langle\tilde{x_j}(n)^2\rangle}},
\end{align}
with averaged cross-correlation
\begin{align}
\label{eq:Gamma}
    \Gamma = \frac{1}{6}{(\Gamma_{12}+\Gamma_{13}+\Gamma_{14}+\Gamma_{23}+\Gamma_{24}+\Gamma_{34})}.
\end{align}
The average is calculated over time after the transients have been discarded. Note that here $\tilde{x}_i(n) = x_i(n) - \langle x_i(n) \rangle$ is the variation of the action potential at index $i$ from its mean. The usual notation for the average over time is given by the angular brackets $\langle \rangle$. The value $\Gamma_{ij}$ can be used to characterize the regimes where the collective dynamics lie. When $\Gamma_{ij} \approx 1$, this means the nodes $i$ and $j$ are almost synchronized with $\Gamma_{ij}=1$ meaning full in-phase synchrony. However, $\Gamma_{ij} \approx -1$ also, meaning the nodes are approaching an anti-phase synchrony, with $-1$ indicating a full anti-phase synchrony. Any value in the range $(-1, 1)$ indicates asynchrony with $\Gamma=0$ meaning complete incoherence/asynchrony. The other values imply partial synchronization behavior, for example {\em chimera} or {\em cluster states} which are more notable when we consider networks in the thermodynamic limits, i.e., the total number of nodes is very high (close to infinity).

\begin{figure}[h]
\centering
\begin{tabular}{cc}
  \includegraphics[scale=0.2]{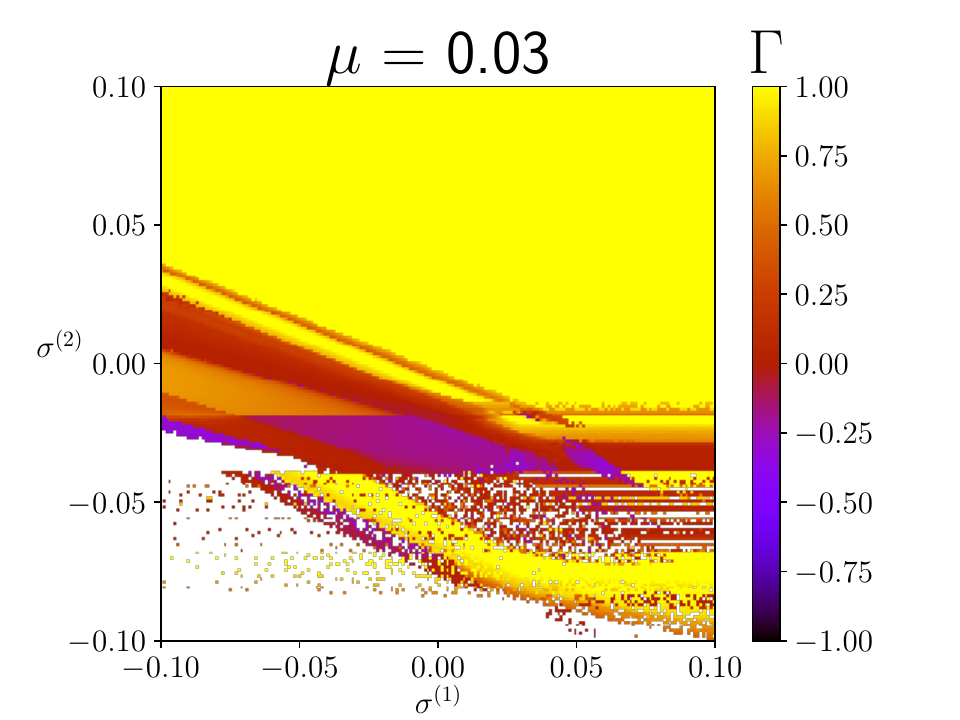} &   \includegraphics[scale=0.2]{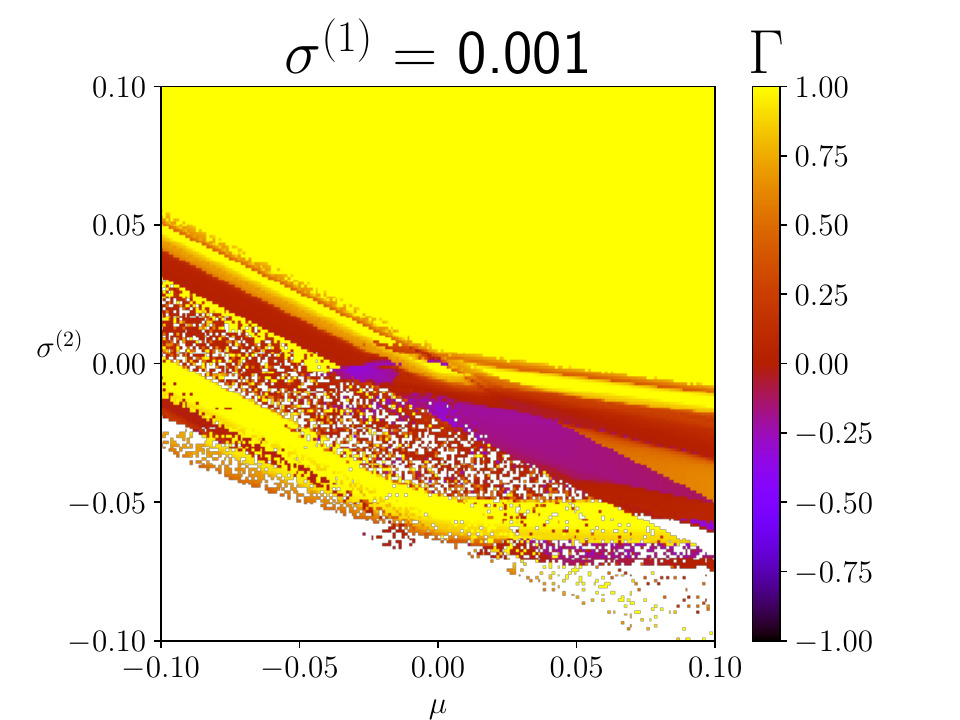} \\
(a) & (b) \\[3pt]
\end{tabular}
\caption{Plots for cross-correlation coefficient. The system is simulated for $5 \times 10^4$ times, and discarding the first $3 \times 10^4$ to ensure transients are removed. Figure (a) represents plot by varying $\sigma^{(1)}$ and $\sigma^{(2)}$ where $\mu = 0.03$, and (b) represents the plot by varying $\mu$ and $\sigma^{(2)}$ where $\sigma^{(1)} =0.001$. Other parameter values are set as $a=0.89, b=0.18, c=0.28$, and $k_0=0.06$. The initial conditions for the plots are sampled uniformly from $[0.6, 0.8]$.}
\label{fig:cross_corln}
\end{figure}

While running the numerics, the simulation was iterated for $5 \times 10^4$ steps, out of which the first $3 \times 10^4$ were discarded. The local parameters are set as $a = 0.89$, $b=0.18$, $c=0.28$, and $k_0 =0.06$. The initial values for the dynamical variables were uniformly sampled from $[0.6, 0.8]$, see Fig.~\ref{fig:cross_corln}. Special attention has been paid to the variation of $\Gamma$ with $\sigma^{(2)}$. Note that in panel (a), we have fixed $\mu = 0.03$ and varied both $\sigma^{(1)}$ and $\sigma^{(2)}$ in the range $[-0.1, 0.1]$, giving rise to a color-coded two-dimensional bifurcation plot. Note that as $\sigma^{(2)}>0$ and becomes more positive, the model falls into a complete synchronization regime. All the nodes oscillate in tandem, as implied by $\Gamma = 1$. When $\sigma^{(2)}<0$, we see that the model usually has $\Gamma<1$, and most of the times ranges between $-0.5$ and $0.5$, meaning highly asynchronous. Further below, as $\sigma^{(2)}$ approaches $-0.1$, the dynamics of the model diverge (the destruction of the attractors) as indicated by the white pixels in the model. In panel (b) we have fixed $\sigma^{(1)}=0.001$ and varied both $\sigma^{(2)}$ and $\mu$ in the range $[-0.1, 0.1]$. Here too we see that when $\sigma^{(2)}$ becomes more positive, the model tends to fall into a complete synchrony regime, whereas when $\sigma^{(2)}$ becomes more negative, the model becomes asynchronous first before seeing a divergence in the behavior as $\sigma^{(2)}$ approaches $-0.1$.

\subsection{Kuramoto order parameter}
The Kuramoto order parameter was first introduced to study the coherence in terms of the phase of the Kuramoto oscillators. Since then, it has found its well-deserved place in the synchronization literature, widely used as an important tool to measure synchrony in agents. The first step is to define the phase $\theta_m$ of an oscillator $m$ at time step $n$. This is given by,
\begin{align}
\label{eq:theta}
\theta_m(n) = \tan^{-1}\left(\frac{y_m{(n)}}{x_m{(n)}}\right).
\end{align}
Next, the complex-valued Kuramoto Index, $I$ can be defined as,
\begin{align}
\label{eq:Kuram}
I_m{(n)} = \exp^{i\theta_m(n)}
\end{align}
where $i=\sqrt{-1}$. Note that our model has four oscillators all coupled to each other. Thus, specific to our model, the index $I(n)$ at time step $n$ is given by,
\begin{align}
   I(n) = \left|\frac{1}{4}\sum_{m=1}^{4}I_m{(n)}\right|.
\end{align}
The notation $|\cdot|$ represents the mean of all phases of the four oscillators inside the unit circle. Thus, the index average over all time is given by,
\begin{align}
\label{eq:Kuram2}
I = \langle{I(n)}\rangle,
\end{align}
which is a time average. Now, if $I =0$, this means the system falls into a total asynchronous regime, whereas $I= 1$ indicates complete asynchrony. Whereas any value $I \in (0, 1)$ indicates partial asynchrony with varying intensity. 

\begin{figure}[h]
\begin{tabular}{cc}
  \includegraphics[scale=0.2]{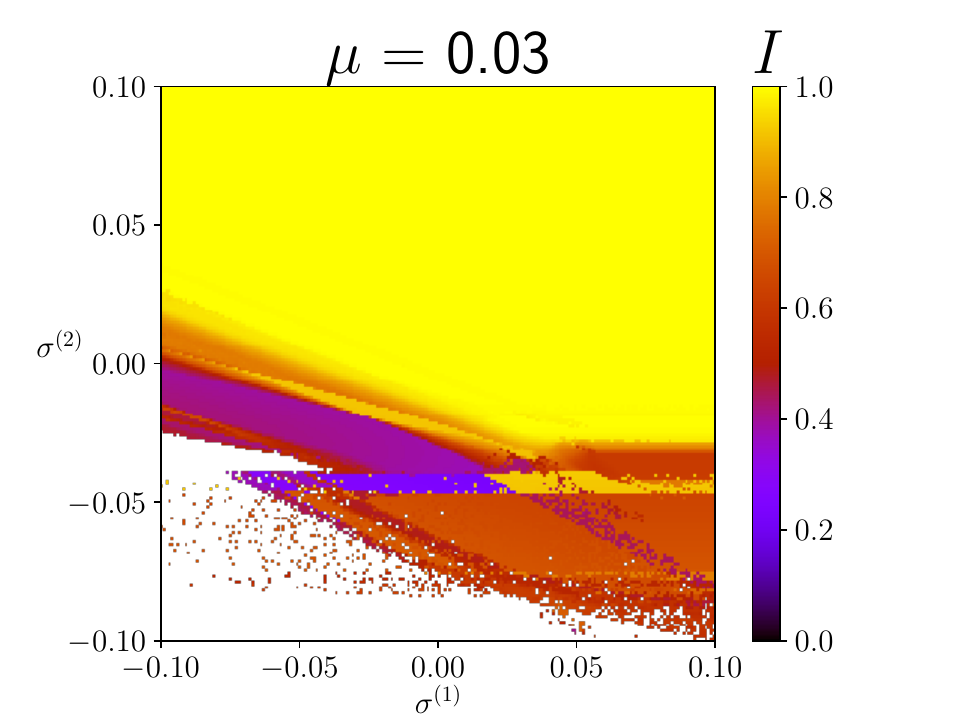} &   \includegraphics[scale=0.2]{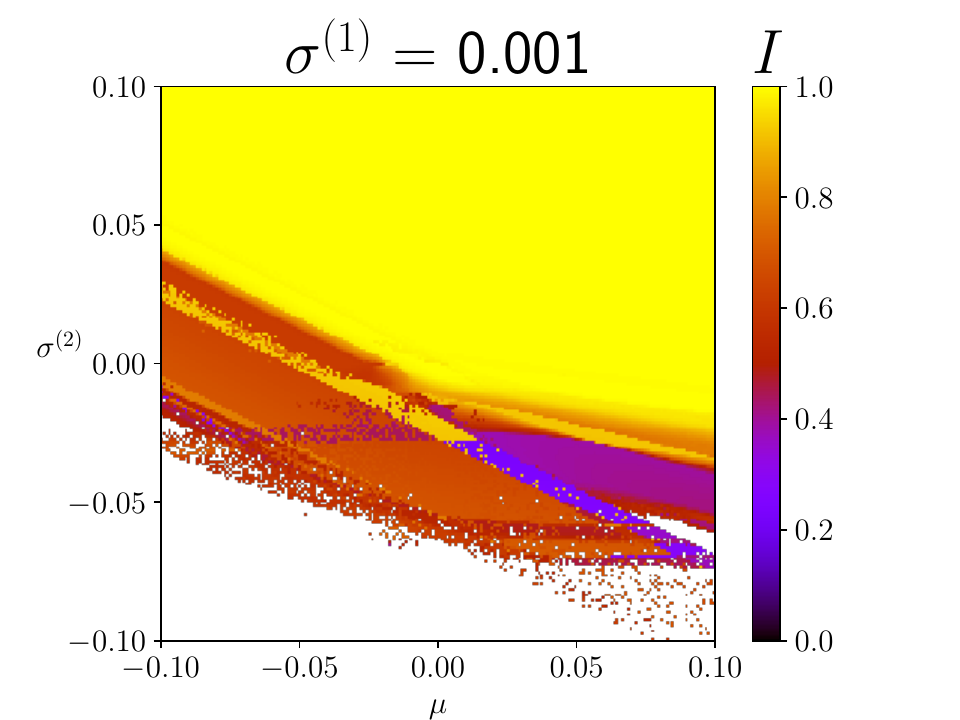} \\
(a) & (b) \\[3pt]
\end{tabular}
    \caption{Plots for Kuramoto-Order Parameter. The system is simulated for $5 \times 10^4$ iterations, and discarding the first $3 \times 10^4$ as transients. Figure (a) represents the plot by varying $\sigma^{(1)}$ and $\sigma^{(2)}$ where $\mu = 0.03$.(b) represents the plot by varying $\mu$ and $\sigma^{(2)}$ where $\sigma^{(1)} =0.001$. Other parameter values are set as $a=0.89, b=0.18, c=0.28, k_0=0.06$. The initial conditions for the plots are sampled uniformly from $[0.6, 0.8]$.}
    \label{fig:kop}
\end{figure}

Similar to Fig.~\ref{fig:cross_corln}, we have also illustrated the two-dimensional color-coded plots for the Kuramoto order parameter, see Fig.~\ref{fig:kop}. Following the same simulation steps as in the cross-correlation coefficient, we have varied $\sigma^{(1)}$ against $\sigma^{(2)}$ in panel (a) in the range $[-0.1, 0.1]$ with $\mu = 0.03$, and varied $\mu$ against $\sigma^{(2)}$ in panel (b) in the range $[-0.1, 0.1]$ with $\sigma^{(1)}=0.001$. We again see in panel (a) that as $\sigma^{(2)}$ becomes more positive, the dynamics of our network become completely synchronized, indicated by $I = 1$ (yellow pixels). As $\sigma^{(2)}$ becomes more negative, however, the synchronization breaks down, with the oscillators oscillating away from one another. This is indicated by $I$ decreasing to a range $[0.23, 0.79]$ approximately. Further below, as $\sigma^{(2)}$ approaches $-0.1$, the network diverges, as illustrated by the white pixels. In panel (b), we see a similar behavior as $\sigma^{(2)}$ is varied. This gives us an overall picture of how the cross-correlation coefficient and the Kuramoto order parameter simulations agree with each other and how they capture the synchronization phenomena in our network. It can be easily seen that the higher order coupling strength significantly drives the dynamics of our network.

\section{Conclusion}
\label{sec:conc}

In this study, we have put forward a novel small network of Chialvo neurons arranged in a ring-star topology made of four neurons (one in the center and three in periphery), where the interactions are modeled beyond pairwise, i.e., higher-order couplings besides pairwise couplings. Furthermore, the couplings are linear and diffusive, replicating an electrical synapse in a realistic nervous system. One future direction is to implement nonlinear couplings that replicate more complicated electrochemical synapses among the neurons. 

Our small network model can be treated as a unit of neuron ensemble which acts as a bridge between a single neuron and a larger ensemble of neurons in the thermodynamic limit (with large numbers of nodes) arranged in a complicated topology coupled linearly or nonlinearly via electrical or chemical synapses, replicating complex functionalities in the nervous system. One interesting avenue to look into in the future is to incorporate noise modulation into the system via additive and multiplicative noises and study a broad range of spatiotemporal patterns. 

This model is a nonlinear system of eight coupled equations which allows us to explore a plethora of dynamical properties via studying its fixed point, performing stability analysis of the fixed point by looking into the eigenvalues of the Jacobian matrix, and reporting bifurcation patterns. The higher-order coupling strength is considered as the primary bifurcation parameter in most cases, and we notice the appearance of an intriguing route to chaos via the appearance of fixed point, period-doubling, cyclic quasiperiodic closed invariant curves, and ultimately chaos. We also observe the emergence of other codimension-$1$ patterns like saddle-node, and Neimark-Sacker. It would be interesting to test what kind of new bifurcation behavior comes to the surface once we consider our network in the thermodynamic limit, and also look into the phenomenon of phase transitions in depth~\cite{GhVe23}.

Additionally, we have looked into quantifying the concept of chaoticity and complexity of the network, by implementing the $0-1$ test and the sample entropy measure on the time series data generated from our model. It would be alluring to look into how a test for chaos and complexity works for a temporal higher-order network where the coupling strengths (pairwise and non-pairwise) update over every iteration. It would also be a good research question to come up with some sort of closed-form analytical relationship between chaos and complexity.

Moreover, we have performed a bifurcation study of synchronization patterns via two-dimensional color-coded plots, measuring how collectively the network behaves over time using the cross-correlation coefficient and the Kuramoto order parameter. We have noticed full synchrony when the higher-order coupling strength becomes more positive, meaning higher-order interaction promotes a tandem cumulative behavior.

In the future, another exercise would be to explore whether these map-based network models have any conservative properties like Hamiltonian in the case of their continuous-time counterparts. Is there an equivalent conserved quantity that we can model for these systems?

Like any other models, we believe our model is not limitation-free and will leave a lot of room for improvement. One way would be to fit our model using empirical data from viable sources. The model can be improved by implementing more complexities for example, time-varying couplings, couplings that are nonlinear, and couplings that integrate $s$-simplex with the highest possible values of $s$. Topologically, it can be improved by modeling a multiplex network, with the first step being a two-layered model and also increasing the number of nodes. The model can be made stochastic by incorporating noise modulations. Studies on finding chimera states, solitary states, traveling waves, and other phenomena like amplitude death could be a future prospect for these small network models.

Our outlook in this paper has been on studying a complex network neurodynamical model via designing the smallest ring-star network. We have come to the conclusion that a high value of the higher-order coupling strength in the excitatory regime will promote more chaoticity and complexity in the system while maintaining full synchrony among the neurons. We believe this will lay a foundation for applications of reduced modeling techniques in the fields of engineering, mathematical modeling, quantitative biology, neuroscience, etc.

\begin{acknowledgments}
A.S.N and S.S.M acknowledge the computing resources provided by Digital University Kerala.
\end{acknowledgments}

\section*{Author Declarations}
The authors have no conflicts to disclose.

\section*{Data Availability Statement}
Data sharing is not applicable to this article as no new data were created or analyzed in this study.

\nocite{*}
\bibliographystyle{unsrt}
\bibliography{main}

\end{document}